\documentclass[12pt]{article}

\usepackage{times}
\usepackage{fullpage}

\usepackage{amsfonts}
\usepackage{amssymb}
\usepackage{amsmath}
\usepackage{latexsym}
\usepackage{url}
\usepackage{hyperref}

\newcommand{\ADV}{\mathrm{ADV}}
\newcommand{\MADV}{\mathrm{ADV}^{\pm}}
\newcommand{\BADV}{\mathrm{ADV}^{(\pm)}}
\newcommand{\AND}{\mathrm{AND}}
\newcommand{\advmax}{\substack{\Gamma \ge 0 \\ \Gamma \ne 0}}
\newcommand{\Tr}{\mathrm{Tr}}

\renewcommand{\H}{\mathcal{H}}

\newcommand{\MM}{\mathrm{MM}}
\newcommand{\fxnefy}[1]{{x,y \atop #1(x) \ne #1(y)}}
\newcommand{\xineyi}[1]{{#1: x_{#1} \ne y_{#1}}}
\newcommand{\gram}{\mathrm{Gram}}
\newcommand{\I}{\mathsf{I}}
\renewcommand{\O}{\mathsf{O}}
\newcommand{\U}{\mathsf{U}}
\newcommand{\id}{\mathrm{id}}

\newcommand{\Gf}{{\Gamma_{\!f}}}

\newcommand{\df}{{\delta_{f}}}
\newcommand{\Gg}[1]{\Gamma_{\!g_{#1}}}
\newcommand{\dg}[1]{{\delta_{g_{#1}}}}
\newcommand{\dgb}[2]{{\delta_{g_{#1}}^{\upharpoonright #2}}}
\newcommand{\dgc}[2]{{\delta_{c_{#1}}^{\upharpoonright #2}}}
\newcommand{\Gh}{{\Gamma_{\!h}}}
\newcommand{\tx}{{\tilde x}}
\newcommand{\ty}{{\tilde y}}
\newcommand{\sublam}[2]{{\|#1 \circ D_{#2}\|}}

\long\def\rem#1{}
\def\01{\{0,1\}}
\newcommand{\bra}[1]{\langle#1|}
\newcommand{\ket}[1]{|#1\rangle}
\newcommand{\braket}[2]{\langle#1|#2\rangle}
\newcommand{\ketbra}[2]{|#1\rangle\langle#2|}
\newcommand{\scalar}[2]{\langle#1,#2\rangle}
\newcommand{\hs}[2]{\langle#1,#2\rangle}
\newcommand{\norm}[1]{\| #1 \|}
\newcommand{\Norm}[1]{\Big\|#1\Big\|}
\newcommand{\sn}[1]{\| #1 \|}
\newcommand{\fro}[1]{\| #1 \|_F}
\newcommand{\trn}[1]{\| #1 \|_{tr}}
\newcommand{\nth}[1]{{\ensuremath{{#1}^{\text{th}}}}}
\newcommand{\nst}[1]{{\ensuremath{{#1}^{\text{st}}}}}
\newcommand{\eps}{\varepsilon}

\newcommand{\R}{\mathbb{R}}
\newcommand{\C}{\mathbb{C}}

\newtheorem{definition}{Definition}

\newtheorem{theorem}{Theorem}
\newtheorem{lemma}[theorem]{Lemma}
\newtheorem{corollary}[theorem]{Corollary}

\newcommand{\thmref}[1]{\hyperref[#1]{{Theorem~\ref*{#1}}}}
\newcommand{\lemref}[1]{\hyperref[#1]{{Lemma~\ref*{#1}}}}
\newcommand{\corref}[1]{\hyperref[#1]{{Corollary~\ref*{#1}}}}
\newcommand{\eqnref}[1]{\hyperref[#1]{{Equation~(\ref*{#1})}}}
\newcommand{\claimref}[1]{\hyperref[#1]{{Claim~\ref*{#1}}}}
\newcommand{\remarkref}[1]{\hyperref[#1]{{Remark~\ref*{#1}}}}
\newcommand{\propref}[1]{\hyperref[#1]{{Proposition~\ref*{#1}}}}
\newcommand{\factref}[1]{\hyperref[#1]{{Fact~\ref*{#1}}}}
\newcommand{\defref}[1]{\hyperref[#1]{{Definition~\ref*{#1}}}}
\newcommand{\exampleref}[1]{\hyperref[#1]{{Example~\ref*{#1}}}}
\newcommand{\hypref}[1]{\hyperref[#1]{{Hypothesis~\ref*{#1}}}}
\newcommand{\secref}[1]{\hyperref[#1]{{Section~\ref*{#1}}}}
\newcommand{\chapref}[1]{\hyperref[#1]{{Chapter~\ref*{#1}}}}
\newcommand{\appref}[1]{\hyperref[#1]{{Appendix~\ref*{#1}}}}

\newcommand{\ignore}[1]{}

\newenvironment{proof}[1][Proof.]{
    \par
    \noindent \textbf{#1}
}{
    \unskip
    \nobreak\hfill\penalty50\hskip3pt\hbox{}\nobreak\hfill
    \hbox{$\Box$}\par\bigskip
}

\allowdisplaybreaks[1]

\begin{document}

\title{Negative weights make adversaries stronger}
\author{%
  Peter H\o yer%
  \thanks{Department of Computer Science,
  University of Calgary.
  Supported by Canada's Natural Sciences and Engineering
  Research Council (NSERC),
  the Canadian Institute for Advanced Research (CIAR),
  and The Mathematics of Information Technology and Complex
  Systems (MITACS).}\\
  {\tt hoyer@cpsc.ucalgary.ca}
\and
  Troy Lee%
  \thanks{%
  LRI, Universit\'e Paris-Sud.  Supported by a Rubicon grant from
  the Netherlands Organisation for Scientific Research (NWO)
  and by the European Commission under the Integrated Project Qubit
  Applications (QAP) funded by the IST directorate as Contract Number 015848.
  Part of this work conducted while at CWI, Amsterdam, and while visiting
  the University of Calgary.}\\
  {\tt troyjlee@gmail.com}
\and
  Robert \v Spalek%
  \thanks{%
  University of California, Berkeley.  Supported by NSF Grant CCF-0524837
  and ARO Grant DAAD 19-03-1-0082.  Work conducted in part while at CWI
  and the University of Amsterdam, supported by the European Commission
  under project QAP, IST-015848, and while visiting the University of Calgary.
  }\\
  {\tt spalek@eecs.berkeley.edu}
}
\date{}
\maketitle

\begin{abstract}
The quantum adversary method is one of the most successful techniques for
proving lower bounds on quantum query complexity.  It gives optimal lower
bounds for many problems, has application to classical complexity in
formula size lower bounds, and is versatile with equivalent formulations in
terms of weight schemes, eigenvalues, and Kolmogorov complexity.
All these formulations rely on the principle
that if an algorithm successfully computes a function then, in particular, it
is able to distinguish between inputs which map to different values.

We present a stronger version of the adversary method which goes beyond this
principle to make explicit use of the stronger condition that the algorithm
actually computes the function.  This new method, which we call
$\MADV$, has all the advantages of
the old: it is a lower bound on bounded-error quantum query complexity, its
square is a lower bound on formula size, and it behaves well with respect to
function composition.  Moreover $\MADV$ is always at least as large as the
adversary method $\ADV$, and we show an example of a monotone function for
which $\MADV(f)=\Omega(\ADV(f)^{1.098})$.  We also give examples showing that
$\MADV$ does not face limitations of $\ADV$ like the certificate
complexity barrier and the property testing barrier.
\end{abstract}

\section{Introduction}
Quantum query complexity is a popular model for study as it seems to capture
much of the power of quantum computing---in particular, the search
algorithm of Grover \cite{Gro96} and the period finding routine of
Shor's factoring algorithm \cite{Sho97} can be formulated in
this model---yet is still simple enough that we can often hope to prove tight
lower bounds.  In this model, complexity is measured by the
number of queries made to the input, and other operations are
for free.  For most known quantum algorithms, the time complexity is bigger
than the query complexity by only a polylogarithmic factor.

The two most successful techniques for proving lower bounds on quantum
query complexity are the polynomial method \cite{BBCMW01} and the quantum
adversary method \cite{Amb02}.  The adversary method gives tight lower bounds
for many problems and is quite versatile with formulations in terms of weight
schemes \cite{Amb03,Zha05}, eigenvalues \cite{BSS03}, and Kolmogorov
complexity \cite{LM04}.  \v{S}palek and Szegedy \cite{SS06} show that in fact
all these formulations are equivalent.
All these versions of the adversary method rest on the principle that, if an
algorithm is able to {\em compute} a function $f$, then in particular it is able
to {\em distinguish} inputs which map to different values.  The method
actually bounds the difficulty of this distinguishing task.

We present a stronger version of the adversary method which goes beyond this
principle to essentially make use of the stronger condition that the algorithm
actually computes the function---namely, we make use of the existence of a
measurement which gives the correct answer with high probability from the
final state of the algorithm.  This new method, which we call $\MADV$, is
always at least as large as the adversary bound $\ADV$, and we show an example
of a monotone function $f$ for which $\MADV(f)=\Omega(\ADV(f)^{1.098})$.
Moreover, $\MADV$ possesses all the nice properties of the old adversary
method: it is a lower bound on bounded-error quantum query complexity, its
square is a lower bound on
formula size, and it behaves well with respect to function composition.
Using this last property, and the fact that our bound is larger than the
adversary bound for the base function of Ambainis, we improve the best known
separation between quantum query complexity and polynomial degree giving an
$f$ such that $Q_{\epsilon}(f)=\Omega(\deg(f)^{1.329})$.

The limitations of the adversary method are fairly well understood.  One
limitation is the ``certificate complexity barrier.''  This says
that $\ADV(f) \le \sqrt{C_0(f)C_1(f)}$ for a total function $f$ with Boolean
output \cite{Zha05,SS06}, where $C_0(f)$ is
the certificate complexity of the inputs $x$ which evaluate to zero on $f$, and
$C_1(f)$ is the certificate complexity of inputs which evaluate to one.
This means that for problems like determining if a graph
contains a triangle, or element distinctness, where one of the certificate
complexities is constant, the best bound which can be proven by the adversary
method is $\Omega(\sqrt{N})$.  For triangle finding, the best known upper
bound is $O(N^{13/20})$ \cite{MSS05}, and for element distinctness the
polynomial method is able to prove a tight lower bound of $\Omega(N^{2/3})$
\cite{AS04}.
We show that our new method can break the certificate complexity barrier---we
give an example where $\MADV(f)=\Omega( (C_0(f)C_1(f))^{0.549})$.

Another limitation of the adversary method is the ``property testing barrier.''
For a partial Boolean function $f$ where all zero-inputs have relative
Hamming distance at least $\epsilon$ from all one-inputs, it holds that
$\ADV(f) \le 1/\epsilon$.  A prime example where this limitation applies is
the collision problem of determining if a function is $2$-to-$1$ or
$1$-to-$1$.  Here all zero-inputs have relative Hamming distance at least
$1/2$ from all one inputs and so the best bound provable by the adversary
method is $2$, while the polynomial method is able to prove a tight lower bound
of $\Omega(n^{1/3})$ \cite{AS04}.  We show the property testing barrier does not
apply in this strict sense to $\MADV$, although we do not know of an
asymptotic separation for constant $\epsilon$.

Breaking these barriers opens the possibility that $\MADV$ can prove tight
lower bounds for problems like element distinctness and the collision problem,
and improve the best known $\Omega(\sqrt{N})$ lower bound for triangle finding.

\subsection{Comparison with previous methods}
We now take a closer look at our new method and how it compares with
previous adversary methods.  We use the setting of the spectral formulation of
the adversary method \cite{BSS03}.

Let $f: S \rightarrow \Sigma_O$ be a function, with $S \subseteq
\Sigma_I^n$ the set of inputs.  We assume $\Sigma_I = \{ 0, 1, \dots,
|\Sigma_I|-1 \}$, and call this the input alphabet and $\Sigma_O$ the output
alphabet. Let $\Gamma$ be a Hermitian matrix with rows and columns
labeled by elements of $S$.  We say that $\Gamma$ is an
\emph{adversary matrix for $f$} if $\Gamma[x,y]=0$ whenever
$f(x)=f(y)$. We let $\sn{M}$ denote the spectral norm of the matrix
$M$, and for a real matrix $M$ use $M\ge 0$ to say the entries of
$M$ are nonnegative. We now give the spectral formulation of the
adversary method:
\begin{definition}
\label{def:adv}
\begin{equation*}
\ADV(f) = \max_{\advmax}\
  \frac{\sn{\Gamma}}{\max_i \sn{\Gamma \circ D_i}}.
\end{equation*}
Here the maximum is taken over nonnegative symmetric adversary matrices
$\Gamma$, and $D_i$ is a zero-one matrix where $D_i[x,y]=1$ if
$x_i \ne y_i$ and $D_i[x,y]=0$ otherwise.  $\Gamma \circ D_i$ denotes the
entry-wise (Hadamard) product of $\Gamma$ and $D_i$.
\end{definition}
Let $Q_{\epsilon}(f)$ be the two-sided $\epsilon$-bounded error quantum query
complexity of $f$.  Barnum, Saks, and Szegedy show that the spectral version of the adversary method is a lower bound on $Q_{\epsilon}(f)$:
\begin{theorem}[\cite{BSS03}]
For any function $f$,
$
Q_{\epsilon}(f)\ge \frac{1-2\sqrt{\epsilon(1-\epsilon)}}{2}\ADV(f).
$
\end{theorem}

Note that the definition of $\ADV(f)$ restricts the maximization to
adversary matrices whose entries are all nonnegative and real.
Our new bound removes these restrictions:
\begin{definition}
$$
\MADV(f) = \max_{\Gamma \ne 0}\
  \frac{\sn{\Gamma}}{\max_i \sn{\Gamma \circ D_i}}.
$$
\end{definition}
It is clear that $\MADV(f) \ge \ADV(f)$ for any function $f$ as the
maximization is taken over a larger set.
Our main theorem, presented in \secref{sec:qqc}, states that $\MADV(f)$ is a
lower bound on $Q_{\epsilon}(f)$.
\begin{theorem}
For any function $f$,
$Q_{\epsilon}(f) \ge \frac{1-2\sqrt{\epsilon(1-\epsilon)}-2\epsilon}{2}
\MADV(f).$
If $f$ has Boolean output, i.e. if $|\Sigma_O|=2$, then
$Q_{\epsilon}(f) \ge \frac{1-2\sqrt{\epsilon(1-\epsilon)}}{2}
\MADV(f).$
\label{thm:qqc}
\end{theorem}

While it is clear that $\MADV$ is always least as large as $\ADV$, it might
at first seem surprising that $\MADV$ can achieve bounds super-linear in $\ADV$.
An intuition for why negative weights help is that it is good to give negative
weight to entries with large Hamming distance, entries which are easier to
distinguish by queries.  Consider an entry $(x,y)$ where
$x$ and $y$ have large Hamming distance.  This entry appears in several
$\Gamma \circ D_i$ matrices but only appears in the $\Gamma$ matrix once.
Thus by giving this entry negative weight we can
simultaneously decrease $\sn{\Gamma \circ D_i}$ for several $i$'s, while
doing relatively little damage to the large $\Gamma$ matrix.

While in form the $\MADV$ bound is very similar to the $\ADV$ bound, our
proof of \thmref{thm:qqc} departs from the standard adversary
principle.  The standard adversary principle is based on the fact that an
algorithm $A$ which is able to \emph{compute} a function $f$ is, in particular,
able to \emph{distinguish} inputs $x,y$ such that $f(x)\ne f(y)$.
Distinguishing quantum states is closely related to the inner
product of the states as given by the following quantitative principle:

\begin{theorem}
Suppose we are given one of two known states $\ket{\psi_x},\ket{\psi_y}$.
Let $0 \le \epsilon \le 1/2$.  There is a measurement which correctly identifies
which of the two states we are given with error probability $\epsilon$ if and
only if $\braket{\psi_x}{\psi_y} \le 2\sqrt{\epsilon(1-\epsilon)}$.
\label{thm:distinguish}
\end{theorem}

Let $\ket{\psi_x^t}$ be the state of an algorithm on input $x$ after $t$ queries.
The adversary method works by defining a ``progress function'' based on the inner
product $\braket{\psi_x^t}{\psi_y^t}$.
Initially, before the algorithm has made any queries, all inputs look the same and
thus $\braket{\psi_x^0}{\psi_y^0}=1$ for all $x,y$, and thus the progress function
is large.  On the other hand, if a $T$-query algorithm computes a function $f$
within error $\epsilon$, then by \thmref{thm:distinguish} for $x,y$ with
$f(x) \ne f(y)$ we must have
$\braket{\psi_x^T}{\psi_y^T} \le 2\sqrt{\epsilon(1-\epsilon)}$, and thus the final
progress function is small.  In \cite{BSS03} this is termed the Ambainis output
condition.  The adversary method then works by showing an upper bound on how much the
progress function can change by a single query.

Our proof follows the same basic reasoning, but the Ambainis output
condition no longer seems to suffice to show that the final progress
function is small. We use in an essential way the stronger output
condition that if a $T$-query algorithm $A$ computes a function $f$,
then there exists orthogonal projectors $\{\Pi_b\}_{b \in \Sigma_O}$
which sum to the identity such that $\|\Pi_b \ket{\psi_x^T}\|^2 \ge
1-\epsilon$ when $f(x)=b$.

\section{Preliminaries}
We assume standard background from quantum computing and Boolean function
complexity, see \cite{NC00} and \cite{BW02} for nice references.  In this
section, we restrict ourselves to more specific background.

\subsection{Linear algebra}
The background we need about matrices can be found in, for example,
\cite{Bha97}.
We use standard notations such as $|\cdot |$ for absolute value,
$\overline{A}$ for the entrywise complex conjugate of a matrix $A$,
$A^*$ for the conjugate transpose of $A$, and $\norm{x}=\sqrt{x^*x}$ for the
$\ell_2$-norm of a vector $x$.
For two matrices $A,B$ of the same size, the Hadamard product or entrywise
product is the matrix $(A \circ B)[x,y]=A[x,y] B[x,y]$.

For an indexed set of vectors $\{\ket{\psi_x}:x \in S\}$, we associate an
$|S|$-by-$|S|$ Gram matrix $M=\gram(\ket{\psi_x}:s \in S)$ where
$$
M[x,y]=\braket{\psi_x}{\psi_y}.
$$
It is easy to see that $M$ is Hermitian and positive semidefinite.

We will make use of several matrix norms.  For a matrix $A$ let $\sn{A}$ be the
spectral norm of $A$
$$
\sn{A}= \max_{x,y} \frac{|x^* A y|}{\norm{x}\norm{y}}.
$$
For two matrices $A,B$ let $\hs{A}{B}$ be the Hilbert-Schmidt inner product.
This is the inner product of $A,B$ viewed as long vectors,
$$
\hs{A}{B}=\Tr(A^*B)=\sum_{i,j} \overline{A[i,j]}B[i,j].
$$
The Frobenius norm, denoted $\fro{A}$, is the norm associated with this inner
product,
$$
\fro{A}=\sqrt{\hs{A}{A}}=\sqrt{\sum_{i,j} |A[i,j]|^2}.
$$
Finally, we will use the trace norm, denoted $\trn{A}$, where
$$
\trn{A}=\max_B \frac{|\hs{A}{B}|}{\sn{B}},
$$
and $B$ runs over all complex matrices of the same size as $A$.
The following theorem is an easy consequence of this definition.
\begin{theorem}
Let $A,B$ be $n$-by-$n$ matrices.  Then
$|\hs{A}{B}| \le \sn{A} \cdot \trn{B}$.
\label{thm:neumann}
\end{theorem}

In our proof that $\MADV$ is a lower bound on quantum query complexity we
will need one more tool for bounding norms:

\begin{theorem}
{\sc (H\"{o}lder's Inequality, \cite{Bha97} Corollary IV.2.6)}
\label{thm:trn}
Let $A,B$ be matrices such that $AB^*$ is defined.  Then
$$
\trn{AB^*} \le \fro{A}\fro{B}.
$$
\end{theorem}

A \emph{partial trace} is a linear mapping $\Tr_A: \mathcal L(A
\otimes B) \to \mathcal L(A)$ mapping linear operators (e.g.,
density matrices) over the joint system $AB$ to linear operators
over $A$.  This mapping is uniquely determined by the requirement
\[
\Tr_A(\rho_A \otimes \rho_B) = \rho_A \cdot \Tr(\rho_B) .
\]
and linearity.
For example, since $\Tr_A(\ketbra i j_A \otimes \ketbra {\psi_i} {\psi_j}_B)
= \ketbra i j_A \cdot \braket {\psi_j} {\psi_i}$, by linearity
\[
\Tr_A \Big( \sum_{i,j} \ketbra i j_A \otimes \ketbra {\psi_i} {\psi_j}_B
\Big) = \sum_{i,j} \ketbra i j_A \cdot \braket {\psi_j} {\psi_i} .
\]

\subsection{Quantum query complexity}

As with the classical model of decision trees, in the quantum query model
we wish to compute some function $f$ and we access the input through queries.
The complexity of $f$ is the number of queries needed to compute $f$ on
a worst-case input $x$.  Unlike the classical case, however, we can
now make queries in superposition.

The memory of a quantum query algorithm is described by three
registers: the input register, $\H_I$, which holds the input $x \in
S \subseteq \Sigma_I^n$, the query register, $\H_Q$, which holds two
integers $1 \le i \le n$ and $0 \le p < |\Sigma_I|$, and the working
memory, $\H_W$, which holds an arbitrary value. The query register
and working memory together form the accessible memory, denoted
$\H_A$.

The accessible memory of a quantum query algorithm $A$ is initialized to
a fixed state.  For convenience, on input $x$ we assume the state
of the algorithm is $\ket x_I \ket{1,0}_Q \ket 0_W$ where all
qubits in the working memory are initialized to 0.
The state of the algorithm then evolves through queries, which depend on the
input register, and accessible memory operators which do not.
We now describe these operations.

We will model a query by a unitary operator where the oracle answer
is given in the phase.  This operator $\O$ is defined by its action
on the basis state $\ket{x} \ket{i,p}$ as
$$
\O\ket{x}\ket{i,p}=e^{\frac {2 \pi \mathbf i} {|\Sigma_I|} p x_i}
\ket{x}\ket{i,p},
$$
where $1 \le i \le n$ is the index of the queried input variable and
$0 \le p < |\Sigma_I|$ is the phase multiplier.  This operation can be
extended to act on the whole space by interpreting it as $\O \times \I_W$,
where $\I_W$ is the identity operation on the workspace $\H_W$.  In the sequel,
we will refer to the action of $\O$ both on $\H_I \otimes \H_Q$ and the full
space $\H_I \otimes \H_Q \otimes \H_W$, and let context dictate which we mean.

For a function with $\Sigma_I=\01$, the query operator simply becomes
$$
\O\ket{x}\ket{i,p}=(-1)^{p x_i} \ket{x}\ket{i,p},
$$

An alternative, perhaps more common, way to model a quantum query is through
an operator $\O': \ket x \ket{i,p} \to \ket x
\ket{i, (x_i + p) \mod |\Sigma_I|}$ that encodes the
result in a register.  These two query models are equivalent, as can
be seen by conjugating with the quantum Fourier transform on $\ket p$.
For our results, it is more convenient to work with the phase oracle.

An accessible memory operator is an arbitrary unitary operation $\U$ on the
accessible memory $\H_A$.  This operation is extended to act on the whole space
by interpreting it as $\I_I \otimes \U$, where $\I_I$ is the
identity operation on the input space $\H_I$.
Thus the state of the algorithm on input $x$ after $t$ queries can be written
$$
\ket{\phi_x^t}= \U_t \O \U_{t-1} \cdots \U_1 \O \U_0 \ket x \ket{1,0} \ket 0.
$$
As the input register is left unchanged by the algorithm, we can decompose
$\ket{\phi_x^t}$ as $\ket{\phi_x^t}=\ket{x}\ket{\psi_x^t}$, where
$\ket{\psi_x^t}$ is the state of the accessible memory after $t$ queries.

The output of a $T$-query algorithm $A$ on input $x$ is chosen
according to a probability distribution which depends on the final
state of the accessible memory $\ket{\psi_x^T}$.  Namely, the
probability that the algorithm outputs some $b \in \Sigma_O$ on
input $x$ is $\|\Pi_b \ket{\psi_x^T}\|^2$, for a fixed set of
projectors $\{\Pi_b\}$ which are orthogonal and complete, that is,
sum to the identity.  The $\epsilon$-error quantum query complexity
of a function $f$, denoted $Q_{\epsilon}(f)$, is the minimum number
of queries made by an algorithm which outputs $f(x)$ with
probability at least $1-\epsilon$ for every $x$.

\section{Bounded-error quantum query complexity}
\label{sec:qqc}
We now show that $\MADV(f)$ is a lower bound on the bounded-error quantum query
complexity of $f$.
\begin{proof}[Proof of \thmref{thm:qqc}]
Let $f:S \rightarrow \Sigma_O$, where $S \subseteq \Sigma_I^n$, be a
function and let $\Gamma$ be a $|S|$-by-$|S|$ Hermitian matrix such
that $\Gamma[x,y]=0$ if $f(x)=f(y)$.  As $\Gamma$ is Hermitian, its eigenvalues
will be equal to its singular values, up to sign.  Notice that
$\Gamma$ and $-\Gamma$ have the same adversary value, thus without loss
of generality we will assume $\Gamma$ has
largest eigenvalue equal to its spectral norm.  Therefore, let $\delta$ be an
eigenvector of $\Gamma$ corresponding to the eigenvalue $\sn{\Gamma}$.

We imagine that we initially prepare the state
$\ket{\Psi^0}=\sum_x \delta_x \ket{x}\ket{1,0}\ket{0}$ and run the algorithm
on this superposition.  Thus after $t$ queries we have the state
\[
\ket{\Psi^t}=\U_t\O\U_{t-1} \ldots \U_1 \O \U_0\sum_x\delta_x\ket{x}\ket{1,0}\ket{0}
=\sum_x \delta_x \ket{x}\ket{\psi_x^t},
\]
where $\ket{\psi_x^t}$ is the state of the accesible memory of the algorithm on
input $x$ after $t$ queries.  We define $\rho^{(t)} = \Tr_I \ket{\Psi^t} \bra{\Psi^t}$ to be the reduced density
matrix of the state $\ket{\Psi^t}$ on the input register, that is we trace out the accessible memory.
In other words, $\rho^{(t)} =\gram(\delta_x \ket{\psi_x^t}: x \in S)$.

We define a progress function $W^t$ based on $\rho^{(t)}$ as
$W^t=\hs{\Gamma}{\rho^{(t)}}$.  Although phrased differently, this
is in fact the same progress function used by H{\o}yer and
\v{S}palek \cite{HS05} in their proof that the regular adversary
method is a lower bound on bounded-error quantum query complexity.
Note that $W^t$ is real, as both $\Gamma$ and $\rho^{(t)}$ are
Hermitian.  Our proof rests on three claims:
\begin{enumerate}
  \item At the beginning of the algorithm $W^0=\sn{\Gamma}$.
  \item With any one query, the progress measure changes by at most
$W^t - W^{t+1} \le 2 \max_i \sn{\Gamma \circ D_i}$.
  \item At the end of the algorithm
    $W^T \le (2\sqrt{\epsilon(1-\epsilon)} + 2\epsilon) \sn{\Gamma}$.
\end{enumerate}

The theorem clearly follows from these three claims.  The main novelty of
the proof lies in the third step.  This is where we depart from the standard
adversary principle in using a stronger output condition implied by a
successful algorithm.

\paragraph*{Item 1}
As the state of the accessible memory $\ket{\psi_x^0}$ is independent of the
oracle, $\braket{\psi_x^0}{\psi_y^0}=1$ for every $x,y$, and so
$\rho^{(0)}=\delta \delta^*$.  Thus
$W^0=\hs{\Gamma}{\delta \delta^*}=\Tr(\delta^* \Gamma^* \delta)=\sn{\Gamma}$.

\paragraph*{Item 2}
After the \nst{t+1} query, the quantum state is $\ket{\Psi^{t+1}} =
\U_{t+1} \O \ket{\Psi^t}$ and thus
\[
\rho^{(t+1)} = \Tr_I (\U_{t+1} \O \ket{\Psi^t} \bra{\Psi^t} \O^*
\U_{t+1}^*) = \Tr_I (\O \ket{\Psi^t} \bra{\Psi^t} \O^*) ,
\]
because the unitary operator $\U_{t+1}$ acts as identity on the
input register. The oracle operator $\O$ only acts on the input
register and the query register, hence we can trace out the working
memory. Denote $\rho = \Tr_{I,Q} \ket{\Psi^t} \bra{\Psi^t}$ and
$\rho' = \O \rho O^*$. Then $\rho^{(t)} = \Tr_I(\rho)$ and
$\rho^{(t+1)} = \Tr_I(\rho')$.  We re-express the progress function
in terms of $\rho, \rho'$. Define two block-diagonal matrices on
$\H_I \otimes \H_Q$:
\begin{align*}
G &= \Gamma \otimes \I_n \otimes \I_{|\Sigma_I|} = \bigoplus
\nolimits_{i=1}^n \bigoplus \nolimits_{p=0}^{|\Sigma_I|-1} \Gamma \\
D &= \bigoplus \nolimits_{i=1}^n \bigoplus
\nolimits_{p=0}^{|\Sigma_I|-1} D_i ,
\end{align*}
where $D_i$ is the zero-one symmetric matrix from \defref{def:adv}.
Then $W^t = \scalar \Gamma {\rho^{(t)}} = \scalar G \rho$ and
$W^{t+1} = \scalar \Gamma {\rho^{(t+1)}} = \scalar G {\rho'}$.  We
upper-bound the change of the progress function as follows:
\begin{align*}
W^t - W^{t+1} &= \scalar G \rho - \scalar G {\rho'} \\
&= \scalar G {\rho - \O \rho \O^*}
    && \rho' = \O \rho \O^* \\
&= \scalar G {(\rho - \O \rho \O^*) \circ D}
    && \rho - \O \rho \O^* = (\rho - \O \rho \O^*) \circ D \\
&= \scalar {G \circ D} {\rho - \O \rho \O^*}
    && \mbox{$D$ is real} \\
&\le \norm{G \circ D} \cdot \trn{\rho - \O \rho \O^*}
    && \mbox{\thmref{thm:neumann}} \\
&\le \norm{G \circ D} \cdot (\trn \rho + \trn{\O \rho \O^*})
    && \mbox{triangle inequality} \\
&= 2 \norm{G \circ D} \cdot \trn \rho
    && \mbox{$\O$ is unitary} \\
&= 2 \norm{G \circ D}
    && \trn \rho = 1 \\
&= 2 \max_i \norm{\Gamma \circ D_i} .
\end{align*}
The third equality holds because $\O$ is diagonal in the
computational basis and thus, in the block corresponding to the
value $\ket{i,p}$ of the query register, $(\rho - \O \rho \O^*)[x,y]
= (1 - e^{\frac {2 \pi \mathbf i} {|\Sigma_I|} p (x_i - y_i)})
\rho[x,y]$, which is $0$ if $x_i = y_i$.  For the remaining
equalities note that conjugating with a unitary operator does not
change the trace norm and that density matrices have trace norm 1.

\paragraph*{Item 3}
Now consider the algorithm at the final time $T$.
We want to upper bound $\hs{\Gamma}{ \rho^{(T)}}$.
The first thing to notice is that as $\Gamma[x,y]=0$ when $f(x)=f(y)$,
we have $\Gamma=\Gamma \circ F$, where $F$ is a zero-one
matrix such that $F[x,y]=1$ if $f(x)\ne f(y)$ and $F[x,y]=0$ otherwise.

As $F$ is a real matrix, it is clear from the definition of the
Hilbert-Schmidt inner product that
$\hs{\Gamma \circ F}{\rho^{(T)}}=\hs{\Gamma}{F \circ \rho^{(T)}}$.
Now applying \thmref{thm:neumann} we have
$\hs{\Gamma}{F \circ \rho^{(T)}} \le \sn{\Gamma} \cdot \trn{\rho^{(T)} \circ F}$.
It remains to upper bound $\trn{\rho^{(T)} \circ F}$, which we do using
\thmref{thm:trn}. To be able to apply this theorem, we would like to find
matrices $X,Y$ such that $XY^*=\rho^{(T)} \circ F$, and the product
$\fro{X}\fro{Y}$ is small.

Let $\{ \Pi_z \}_{z \in \Sigma_O}$ be a complete set of orthogonal
projectors that determine the output probabilities, that is the probability
that the algorithm outputs $z$ on input $x$ is $\norm{\Pi_z \ket{\psi_x^T}}^2$.
We will use these projectors to help decompose $\rho^{(T)}$ in such a way
as to apply \thmref{thm:trn}.
The correctness of the algorithm tells us that
$\norm{\Pi_{f(x)} \ket{\psi_x^T}}^2 \ge 1 - \eps$.  For an $i \in \{
0, 1, \dots, |\Sigma_O|-1 \}$, let $X_i$ denote the matrix with
$|S|$ rows $\{ \Pi_{f(x)+i} \delta_x \ket{\psi_x^T} \}_{x \in S}$, where
$f(x)+i$ is computed modulo $|\Sigma_O|$.
Intuitively, $X_0$ is the matrix where we project onto the correct answers,
and $X_i$ for $i \ge 1$ where we project onto some incorrect answer.
The matrices $X_i$ for $i \ge 1$ will therefore have small Frobenius norm.

We claim $\rho^{(T)} \circ F = \sum_{i \ne j} X_i X_j^*$:
\begin{align*}
\Big( \sum_{i \ne j} X_i X_j^* \Big)[x,y] &= \sum_{i \ne j} \delta_x
\delta_y^* \cdot
    \bra{\psi_y^T} \Pi_{f(y)+j} \Pi_{f(x)+i} \ket{\psi_x^T} \\
&= \begin{cases}
0 & f(x) = f(y) \\
\delta_x \delta_y^* \braket {\psi_y^T} {\psi_x^T} & f(x) \ne f(y)
\end{cases} \\
&= (\rho^{(T)} \circ F)[x,y] ,
\end{align*}
because $\Pi_{z_1} \Pi_{z_2} = 0$ for $z_1 \ne z_2$, $\Pi_z^2 =
\Pi_z$, and $\sum_z \Pi_z = I$. Then
\begin{align*}
\trn{\rho^{(T)} \circ F}
&= \Norm{\sum_{i \ne j} X_i X_j^*}_{tr} \\
&\le \Norm{\sum_{i \ge 1} (X_0 X_i^* + X_i X_0^*)}_{tr}
    + \Norm{\sum_{\substack{i \ne j\\ i,j \ge 1}} X_i X_j^*}_{tr}
    && \mbox{triangle inequality} \\
&= \trn{X_0 (X_0^\perp)^* + X_0^\perp X_0^*}
    + \underbrace{\Norm{\sum_{i,j \ge 1} X_i X_j^* - \sum_{i \ge 1} X_i X_i^*}_{tr}}
    _{\mbox{only present for $|\Sigma_O| > 2$}}
    && \mbox{define } X_0^\perp = \sum_{i \ge 1} X_i \\
&\le 2 \trn{X_0 (X_0^\perp)^*}
    + \trn{X_0^\perp (X_0^\perp)^*}
    + \Norm{\sum_{i \ge 1} X_i X_i^*}_{tr}
    && \mbox{triangle inequality} \\
&\le 2 \fro{X_0} \fro{X_0^\perp}
    + \fro{X_0^\perp}^2 + \Norm{\sum_{i \ge 1} X_i X_i^*}_{tr} .
    && \mbox{\thmref{thm:trn}}
\end{align*}
We now bound the term $\trn{\sum_{i \ge 1} X_i X_i^*}$.  As each $X_iX_i^*$ is
positive semidefinite, the trace norm of this sum is equal to its trace.
Notice that $\Tr(X_i X_j^*) = 0$ if $i \ne j$, because $(X_i X_j^*)[x,x] =
|\delta_x|^2 \bra {\psi_x^T} \Pi_{f(x)+i} \Pi_{f(x)+j}
\ket{\psi_x^T} = 0$.  Thus
\[
\Norm{\sum_{i \ge 1} X_i X_i^*}_{tr}
= \Tr\Big( \sum_{i \ge 1} X_i X_i^* \Big)
= \Tr( X_0^\perp (X_0^\perp)^* )
= \fro{ X_0^\perp }^2 .
\]
Therefore $\trn{\rho^{(T)} \circ F} \le 2 a b + 2 b^2$, where $a =
\fro{X_0}$ and $b = \fro{X_0^\perp}$.  We know that $a,b$ satisfy
the following constraints:
\[
a^2 + b^2
= \fro{X_0}^2+\fro{X_0^\perp}^2
= \sum_{x\in S} |\delta_x|^2 (\|\Pi_{f(x)} \ket{\psi_x^T}\|^2
    + \|(I - \Pi_{f(x)}) \ket{\psi_x^T}\|^2)
= \norm \delta^2
= 1
\]
and
\[
a^2
= \fro{X_0}^2
= \sum_{x \in S} |\delta_x|^2 \|\Pi_{f(x)} \ket{\psi_x^T}\|^2
\ge (1-\epsilon) \sum_{x \in S} |\delta_x|^2
= 1 - \epsilon .
\]
Assuming $\eps \le \frac 1 2$, the maximum of the expression $2 a b + 2 b^2$
under these constraints is the boundary case $a = \sqrt{1 - \eps}$. Hence
$\trn{\rho^{(T)} \circ F} \le 2 \sqrt{\eps (1-\eps)} + 2 \eps$.  As we
note above, the bound can be strengthened to $2 \sqrt{\eps
(1-\eps)}$ if the function has Boolean output, that is $|\Sigma_O|=2$.
\end{proof}

\section{Formula size}
Laplante, Lee, and Szegedy \cite{LLS06} show that
the adversary method can also be used to prove classical lower bounds---they
show that $\ADV(f)^2$ is a lower bound on the formula size of $f$.  A formula
is circuit with AND, OR, and NOT gates with the restriction that every gate
has out-degree exactly one.  The size of a formula is the number of leaves and
the size of a smallest formula computing $f$ is denoted $L(f)$.  We show
that $\MADV(f)^2$ remains a lower bound on the formula size of $f$.

Before we prove this statement, note that it implies a
limitation of $\MADV(f)$---it is upper bounded by the square root of the formula
size of $f$.  Thus for the binary AND-OR tree---or read-once formulae in
general--- the largest lower bounds provable by $\MADV$ are $\sqrt{n}$.
Laplante, Lee, and Szegedy conjecture that this is not a limitation at all---
that is, they conjecture that bounded-error quantum query complexity squared
is in general a lower bound on quantum query complexity.  A major step has
recently been taken toward proving this conjecture by \cite{FGG07,CRSZ07}, who
show that $Q_2(f) \le L(f)^{1/2+\epsilon}$ for any $\epsilon > 0$.

We will work in the setting of Karchmer and Wigderson, who characterize formula
size in terms of a communication complexity game \cite{KW88}.
Since this seminal work, nearly all formula size lower bounds have been
formulated in the language of communication complexity.

Let $f:\{0,1\}^n \rightarrow \{0,1\}$ be a Boolean function.  Following
Karchmer and Wigderson, we associate with $f$ a relation
$R_f \subseteq \01^n \times \01^n \times [n]$ where
$$
R_f=\{(x,y,z): f(x)=0, f(y)=1, x_z \ne y_z\}.
$$
For a relation $R$, let $C^P(R)$ denote the number of leaves in a smallest
communication protocol for $R$, and let $L(f)$ be the number of leaves in a smallest
formula for $f$.  Karchmer and Wigderson show the following:

\begin{theorem}
$L(f)=C^P(R)$.
\end{theorem}

We say that a set $S \subseteq X \times Y$ is \emph{monochromatic} with
respect to
$R$ if there exists $z \in Z$ such that $(x,y,z) \in R$ for all $(x,y) \in S$.
It is well known, see for example \cite{KN97}, that a successful
communication protocol for a relation
$R \subseteq X \times Y \times Z$ partitions $X \times Y$ into disjoint
combinatorial rectangles which are monochromatic with respect to $R$.
Let $C^D(R)$ be the size of a smallest decomposition of $X \times Y$ into
disjoint rectangles monochromatic with respect to $R$.  Clearly,
$C^D(R) \le C^P(R)$.  We are actualy able to show the stronger statement that
the square of $\MADV(f)$ is a lower bound on the size of a smallest rectangle
decomposition of $R_f$.

\begin{theorem}
$L(f) \ge C^D(R_f) \ge (\MADV(f))^2$.
\end{theorem}

\begin{proof}
Laplante, Lee, and Szegedy \cite{LLS06} show that two conditions are
sufficient for a measure to lower bound formula size.  The first is
rectangle subadditivity---they
show that the spectral norm squared is subadditive over rectangles,
and this result holds for an arbitrary, possibly negative, matrix.

\begin{lemma}[Laplante, Lee, Szegedy]
Let $A$ be an arbitrary $|X|$-by-$|Y|$ matrix and $\mathcal R$ a rectangle
partition of $|X| \times |Y|$.  Then
$\sn{A}^2 \le \sum_{R \in \mathcal{R}} \sn{A_R}^2$.
\end{lemma}

The second property is monotonicity, and here we need to modify their argument
to handle negative entries.  They use the property that if $A,B$ are
nonnegative matrices, and if $A \le B$ entrywise, then $\sn{A}\le \sn{B}$.
In our application, however, we actually know more: if $R$ is a
rectangle monochromatic with respect to a color $i$, then $A_R$ is a
\emph{submatrix} of $A_i$.  And, for arbitrary matrices $A,B$, if $A$ is a
submatrix of $B$ then $\sn{A} \le \sn{B}$.

This allows us to complete the proof: let $\mathcal{R}$ be a monochromatic
partition of $R_f$ with $|\mathcal{R}|=C^D(R_f)$.  Then for any matrix $A$
\begin{align*}
\sn{A}^2 &\le \sum_{R \in \mathcal{R}} \sn{A_R}^2 \le
C^D(R_f) \cdot \max_R \sn{A_R}^2 \\
&\le C^D(R_f) \cdot \max_i \sn{A_i}^2.
\end{align*}
And so we conclude
$$
L(f) \ge C^D(R_f) \ge \max_{A \ne 0} \frac{\sn{A}^2}{\max_i \sn{A_i}^2}.
$$
\end{proof}

\section{Automorphism Principle}
In practice, many of the functions we are interested in possess a high degree
of symmetry.  We now show how to take advantage of this symmetry to simplify
the computation of the adversary bound.  We will state this principle in a
general way for possibly non-Boolean functions.  Thus let
$\Sigma,T$ be two finite sets, and let $f:\Sigma^n \rightarrow T$ be a
function.  We will define a group action of $S_n \times S_{\Sigma}^n$
on our set of inputs $\Sigma^n$.  A natural action of an element
$\tau \in S_n$ on input $x \in \Sigma^n$ is to permute the indices of $x$.
Namely, we define $\tau \cdot x=y$ to be the string where $x_i=y_{\tau(i)}$.
An element $\sigma \in S_{\Sigma}^n=(\sigma_1,\ldots, \sigma_n)$ similarly has
a natural action on an input $x \in \Sigma^n$.  Namely, we define
$\sigma \cdot x=y$ to be the string where $y_i=\sigma_i(x_i)$.  In general we
can combine these actions and for
$(\tau,\sigma) \in S_n \times S_{\Sigma}^n$ define
$(\tau,\sigma) \cdot x= \tau \cdot (\sigma \cdot x)$.  From now on, we will
use the more convenient functional notation
$(\tau,\sigma)(x)=\tau \cdot (\sigma \cdot x)$.

\begin{definition}
Let $f:\Sigma^n \rightarrow T$ be a function, and
$\pi \in S_n \times S_{\Sigma}^n$.  We say that $\pi$ is an automorphism
of $f$ if $f(x) \ne f(y) \Rightarrow f(\pi(x)) \ne f(\pi(y))$ for
all $x,y \in \Sigma^n$.
\end{definition}
Note that the automorphisms of a function form a group.

Intuitively, when choosing a weight matrix $\Gamma$, it seems that pairs
$(x,y)$ and $(\pi(x),\pi(y))$ ``look the same'' when $\pi$ is an
automorphism of $f$ and therefore should be given the same weight.
The automorphism principle makes this intuition rigorous.  This principle
can vastly simplify the computation of the adversary bound, helping both in
choosing good weight matrices, and in showing upper bounds on the adversary
value.

\begin{definition}
Let $G$ be a group of automorphisms for a function $f$.  We say that
$G$ is $f$-transitive if for every $x,y$ such that $f(x)=f(y)$, there is
$\pi \in G$ such that $\pi(x)=y$.
\end{definition}

\begin{theorem}[Automorphism Principle]
Let $G$ be a group of automorphisms of $f$.  There is an optimal adversary
matrix $\Gamma$ for which $\Gamma[x,y]=\Gamma[\pi(x),\pi(y)]$ for all
$\pi \in G$ and $x,y$.  Furthermore, if $G$ is $f$-transitive then $\Gamma$
has a principal eigenvector $\beta$ for which $\beta[x]=\beta[y]$ whenever
$f(x)=f(y)$.
\label{thm:auto}
\end{theorem}

\begin{proof}
Let $\Gamma$ be an optimal adversary matrix for $f$.  By normalizing as
necessary, we assume that $\MADV(f)=\sn{\Gamma}$---that is,
$\max_i \sn{\Gamma \circ D_i} = 1$.  For an automorphism
$\pi$, we let $\Gamma_{\pi}$ be the matrix obtained from $\Gamma$ by
permuting the rows and columns by $\pi$, that is
$\Gamma_{\pi}[x,y]=\Gamma[\pi(x),\pi(y)]$.  Letting $P$ be the permutation
matrix representing $\pi$, where $P[x,y]=1$ if $x=\pi(y)$ and 0 otherwise,
we see that $P^T \Gamma P=\Gamma_{\pi}$.  As $P$ is unitary, this means
that $\sn{\Gamma}=\sn{\Gamma_{\pi}}$.  Notice that if
$\pi$ sends the index $i$ to $j$, then if $x_i \ne y_i$ it follows
$\pi(x)_j \ne \pi(y)_j$.  Thus
$P^T (\Gamma \circ D_i) P=\Gamma_{\pi} \circ D_j$, and these matrices also
have the same spectral norm.  It follows that $\Gamma_{\pi}$ achieves
the same adversary bound as $\Gamma$, and so is an optimal adversary matrix.

Let $\delta$ be a principal eigenvector of $\Gamma$.  We may assume without
loss of generality that all entries of $\delta$ are nonzero, as the rows and
columns of $\Gamma$ corresponding to zero entries of $\delta$ can be removed
without affecting the adversary bound.  We now see how $\delta$ relates
to a principal eigenvector of $\Gamma_{\pi}$: note that
$P^T \Gamma P (P^T \delta)=P^T \Gamma PP^T \delta=\norm \Gamma P^T \delta$,
thus $P^T \delta$ will be a principal eigenvector of $\Gamma_{\pi}$. The vector
$P^T \delta$ has entries $P^T \delta[x]=\delta[\pi(x)]$.  For convenience
we set $\delta_{\pi}=P^T \delta$.

To prove the automorphism principle, we will now ``average'' the matrices
$\Gamma_{\pi}$ over $\pi \in G$ in the following way.
We define a vector $\beta$ as
$$
\beta[x]=\sqrt{\sum_{\pi \in G} \delta[\pi(x)]^2}
$$
Notice that $\beta$ has norm $\sqrt{|G|}$.
Now form the matrix $\Gamma'$, where
\begin{equation}
\Gamma'[x,y]=\frac{\sum_{\pi \in G} \Gamma[\pi(x),\pi(y)]
\delta[\pi(x)]^*\delta[\pi(y)]}
{\beta[x]^* \beta[y]}
\label{eq:avg_gamma}
\end{equation}

We claim that $\Gamma'$ is an optimal adversary matrix.  The spectral
norm of $\Gamma'$ is at least
$$
\sn{\Gamma'}\ge \frac{1}{|G|} \beta^* \Gamma' \beta=
\frac{1}{|G|} \sum_{\pi \in G} \sn{\Gamma_{\pi}}=\sn{\Gamma}.
$$

We will now show that $\sn{\Gamma' \circ D_i} \le 1$ for all $i \in [n]$.
This is equivalent to showing $I \pm \Gamma' \circ D_i \succeq 0$.
As argued above, we have $I \pm \Gamma_{\pi} \circ D_i \succeq 0$, for
all $\pi \in G$.  It follows that also
$\delta_{\pi} \delta_{\pi}^* \circ (I \pm \Gamma_{\pi} \circ D_i)
\succeq 0$.  Now adding these equations over $\pi \in G$ we obtain
$$
\left(\sum_{\pi \in G} \delta_{\pi} \delta_{\pi}^* \right) \circ I \pm
\left(\sum_{\pi \in G} \Gamma_{\pi} \circ \delta_{\pi} \delta_{\pi}^* \right)
\circ D_i \succeq 0.
$$
We can further take the Hadamard product of this matrix with the rank
one matrix $A$, where $A[x,y]=[1/\beta[x]^*\beta[y]]$, with the result
remaining positive semidefinite.  This gives
$$
I \pm (\sum_{\pi \in G} \Gamma_{\pi} \circ \delta_{\pi} \delta_{\pi}^*)
\circ D_i \circ A = I \pm \Gamma' \circ D_i \succeq 0,
$$
which concludes the proof that $\Gamma'$ is an optimal adversary matrix.

Note that this argument has actually shown that $\beta$ is a principal
eigenvector of $\Gamma'$.  Let $\sigma$ be an arbitrary element of
$G$.  We have
\begin{equation*}
\beta[\sigma(x)]= \sqrt{\sum_{\pi \in G} \delta[\pi \sigma(x)]^2}
                = \sqrt{\sum_{\pi \sigma^{-1}:\pi \in G} \delta[\pi(x)]^2}
                = \beta[x].
\end{equation*}
If $G$ is $f$-transitive, then for every $x,y$ with $f(x)=f(y)$, there is
some $\sigma$ such that $\sigma(x)=y$, and thus $\beta(x)=\beta(y)$.
This proves the ``furthermore'' of the theorem.

Now we will show that $\Gamma'[x,y]=\Gamma'[\sigma(x),\sigma(y)]$.  We
have just argued that $\beta[x]=\beta[\sigma(x)]$, which gives that the
denominators of these terms, defined by \eqnref{eq:avg_gamma}, are equal.
That the numerators are equal follows similarly since summing over
$\pi \in G$ is the same as summing over $\pi \sigma^{-1}: \pi \in G$ as
$G$ is a group.
\end{proof}

We single out another class of functions which arise frequently in practice
and where the automorphism principle can give the adversary bound a
particularly simple form.

\begin{definition}
Let $f:\Sigma^n \rightarrow T$ be a function, and let
$G \subseteq S_n \times S_{\Sigma}^n$ be its group of automorphisms.  We
say that $G$ is index transitive if for every $i,j \in [n]$ there is an
automorphism $\pi=(\sigma,\tau) \in G$ with $\sigma(i)=j$.
\end{definition}

\begin{corollary}
If $f$ has a group of automorphisms which is index transitive,
then there is an optimal adversary matrix $\Gamma$
such that $\sn{\Gamma \circ D_i}=\sn{\Gamma \circ D_j}$ for all $i,j \in [n]$.
\end{corollary}

This corollary means that if a function has an automorphism group which is
index transitive we can do away with the maximization in the denominator of
the adversary bound---all $\Gamma \circ D_i$ will have the same spectral norm.

\begin{proof}
Let $i,j \in [n]$, and let $G$ be the group of automorphisms
of $f$.  By assumption, there is a $\pi=(\sigma, \tau) \in G$ such that
$\sigma(i)=j$.  Applying \thmref{thm:auto}, there is an optimal adversary
matrix $\Gamma$ such that $\Gamma[\pi(x),\pi(y)]=\Gamma[x,y]$, for all
$x,y$.  Notice that $x_i \ne y_i$ if and only if $\pi(x)_j \ne \pi(y)_j$.
Letting $P$ be the permutation matrix representing $\pi$, it then follows
that $P^T(\Gamma \circ D_i)P=\Gamma \circ D_j$, and so $\Gamma \circ D_i$ and
$\Gamma \circ D_j$ have the same spectral norm.
\end{proof}

\section{Composition theorem}
One nice property of the adversary method is that it behaves very
well with respect to iterated functions.  In this section we will
exclusively deal with Boolean functions.  For a function
$f:\01^n \rightarrow \01$ we define the \nth{d} iteration of $f$,
$f^d:\01^{n^d} \rightarrow \01$ recursively as
$f^1 = f$ and $f^{d} = f \circ (f^{d-1}, \dots, f^{d-1})$ for $d > 1$.
Ambainis~\cite{Amb03} shows that $\ADV(f^d) \geq \ADV(f)^d$.
Thus by proving a good adversary bound on the base function $f$, one can
easily obtain good lower bounds on the iterates of $f$.
In this way, Ambainis shows a super-linear gap between the
bound given by the polynomial degree of a function and the adversary
method, thus separating polynomial degree and quantum query complexity.

Laplante, Lee, and Szegedy~\cite{LLS06} show
a matching upper bound for iterated functions, namely that if $\ADV(f) \leq a$
then $\ADV(f^d) \leq a^d$.
Thus we conclude that the adversary method possesses the following composition
property.
\begin{theorem}[\cite{Amb03,LLS06}]
\label{thm:iterated}
For any function $f:S \rightarrow \01$, with $S \subseteq \01^n$ and natural
number $d>0$,
\begin{equation*}
\ADV(f^d) = \ADV(f)^d.
\end{equation*}
\end{theorem}

H{\o}yer, Lee, and \v{S}palek \cite{HLS05} generalize this composition
theorem to functions that can be written in the form
\begin{equation}\label{eq:composed}
h = f \circ (g_1, \dots, g_k).
\end{equation}
They give an exact expression for the adversary bound of $h$ in terms
of the adversary bounds of $f$ and $g_i$ for $1\le i \le k$.  We will
also look at the composition of the $\MADV$ bound in this general setting.

One may think of $h$ as a two-level decision tree with the top node
being labeled by a function $f:\01^k \rightarrow \01$, and each of
the $k$ internal nodes at the bottom level being labeled by a
function $g_i:\01^{n_i} \rightarrow \01$.  We do not require that the
inputs to the inner functions $g_i$ have the same length.  An~input $x
\in \01^n$ to $h$ is a bit string of length $n = \sum_i n_i$, which we
think of as being comprised of $k$ parts, $x=(x^1, x^2, \ldots, x^k)$,
where $x^i \in \01^{n_i}$.  We may evaluate $h$ on input $x$ by first
computing the $k$ bits $\tx_i = g_i(x^i)$, and then evaluating $f$ on
input $\tx = (\tx_1, \tx_2, \ldots, \tx_k)$.

\paragraph{Adversary bound with costs}
To show their composition theorem, \cite{HLS05} consider as an intermediate
step a generalization of the adversary method allowing input bits to be given
an arbitrary positive cost.  For any function $f: \01^n \rightarrow \01$,
and any vector $\alpha \in \R_{+}^n$ of length $n$ of positive
reals, they define a quantity $\ADV_\alpha(f)$ as follows:
\[
\ADV_\alpha(f) = \max_{\advmax} \min_i \left\{ \alpha_i
  {\sn{\Gamma} \over \sn{\Gamma \circ D_i}} \right\}.
\]
We define the analogous quantity $\MADV_{\alpha}(f)$ by enlarging the
maximization over all nonzero adversary matrices.  We will use the
notation $\BADV$ to simultaneously refer to both $\ADV$ and $\MADV$.
One may think of $\alpha_i$ as expressing the cost of querying the \nth{i}
input bit $x_i$.  For example, $x_i$ could be equal to the parity of
$\alpha_i$ new input bits, or,
alternatively, each query to $x_i$ could reveal only a fraction of
$1/\alpha_i$ bits of information about~$x_i$.  When $\alpha=(a,\ldots,a)$
and all costs are equal to $a$, the new adversary bound $\BADV_\alpha(f)$
reduces to $a \cdot \BADV(f)$, the product of $a$ and the adversary
bound $\BADV(f)$.  In particular, when all costs $a=1$ we have
$Q_{\epsilon}(f)=\Omega(\BADV_{\vec 1}(f))$.
When $\alpha$ is not the all-one vector, then $\BADV_{\alpha}(f)$ will not
necessarily be a lower bound on the quantum query complexity of $f$, but
this quantity can still be very useful in computing the adversary bound of
composed functions.  We will show the following
composition theorem for the nonnegative $\ADV$ bound:

\begin{theorem}
[Exact expression for adversary bound of composed functions]
\label{thm:composition} For any
function $h: S \rightarrow \01$ of the form $h =
f \circ (g_1, \dots, g_k)$ with domain $S \subseteq \01^n$,
and any cost function $\alpha \in \R_{+}^n$,
\begin{equation*}
\ADV_\alpha(h) = \ADV_\beta(f),
\end{equation*}
where $\beta_i = \ADV_{\alpha^i}(g_i)$,
$\alpha = (\alpha^1, \alpha^2, \ldots, \alpha^k)$,
and $\beta = (\beta_1, \ldots, \beta_k)$.
\end{theorem}

We show that one direction of this theorem, the lower bound, also holds for
the $\MADV$ bound.  This is the direction which is useful for proving
separations.
\begin{theorem}
\label{thm:madvcomposition}
Let $h,f,g_i$ be as in the previous theorem.  Then
\begin{equation*}
\MADV_\alpha(h) \ge \MADV_\beta(f),
\end{equation*}
where $\beta_i = \MADV_{\alpha^i}(g_i)$,
$\alpha = (\alpha^1, \alpha^2, \ldots, \alpha^k)$,
and $\beta = (\beta_1, \ldots, \beta_k)$.
\end{theorem}
As with the proof that $\MADV$ is a lower bound on quantum query complexity,
the presence of negative entries again causes new difficulties in the proof
of the composition theorem.  In particular, previous proofs of composition
theorems do not seem to work for $\MADV$ and we prove
\thmref{thm:madvcomposition} in a quite different manner.  Also, as the dual
of the $\MADV$ bound is more complicated than that of the $\ADV$ bound, we
have not yet been able to show the upper bound in this theorem.

The usefulness of such a theorem is that it allows one to divide and
conquer---it reduces the computation of the adversary bound for $h$ into the
disjoint subproblems of first computing the adversary bound for each $g_i$,
and then, having determined $\beta_i=\BADV(g_i)$, computing
$\BADV_\beta(f)$, the adversary bound for $f$ with costs $\beta$.

One need not compute exactly the adversary bound for each $g_i$ to apply the
theorem.  Indeed, a bound of the form $a \le \ADV(g_i) \le b$ for
all $i$ already gives some information about $h$.
\begin{corollary}
If $h=f \circ (g_1, \ldots, g_k)$ and $a \le \ADV(g_i) \le b$ for all $i$,
    then $a \cdot \ADV(f) \le \ADV(h) \le b \cdot \ADV(f)$.
\end{corollary}

One limitation of our theorem is that we require the
sub-functions $g_i$ to act on disjoint subsets of the input bits.  Thus one
cannot use this theorem to compute the adversary bound of any function by,
say, proceeding inductively on the structure of a
$\{\wedge,\vee,\neg\}$-formula
for the function.  One general situation where the theorem can be applied,
however, is to read-once functions, as by definition these functions are
described by a formula over $\{\wedge, \vee, \neg\}$ where each variable
appears only once.

To demonstrate how \thmref{thm:composition} can be
applied, we give a simple proof of the $\Omega(\sqrt{n})$ lower bound due to
Barnum and Saks \cite{BS04} on the bounded-error quantum query complexity of
read-once functions.

%We assume throughout this paper that the
%functions $f$, $g_i$, and $h$ are boolean.  We do not assume that $h$
%is total---if $h = f \circ (g_1, \ldots, g_k)$ is partial, the domain
%of $f$, and of each sub-function $g_i$, is given by the possible
%inputs that may be generated to them via valid inputs to~$h$.
%when there is no subscript, we mean the ordinary adversary bound.

\begin{corollary}[Barnum-Saks]
Let $h$ be a read-once Boolean function with $n$ variables.  Then
$Q_{\epsilon}(h) = \Omega(\sqrt{n})$.
\label{cor:bs}
\end{corollary}

\begin{proof}
We prove by induction on the number of variables $n$ that
$\ADV(f) \ge \sqrt{n}$.  If $n=1$ then the formula is either $x_i$ or
$\neg x_i$ and taking $\Gamma=1$ shows the adversary bound is at least 1.

Now assume the induction hypothesis holds for read-once formulas on
$n$ variables, and let $h$ be given by a read-once formula with $n+1$ variables.
As usual, we can push any NOT gates down to the leaves, and assume that the
root gate in the formula for $h$ is labeled either by an AND gate or an OR gate.
Assume it is AND---the other case follows similarly.  In this case, $h$ can be
written as
$h=g_1 \wedge g_2$ where $g_1$ is a read-once function on $n_1 \le n$ bits
and $g_2$ is a read-once function on $n_2 \le n$ bits, where $n_1+n_2=n+1$.
We want to calculate $\ADV_{\vec 1}(h)$.  Applying \thmref{thm:composition},
we proceed to first calculate $\beta_1=\ADV(g_1)$ and $\beta_2=\ADV(g_2)$.
By the induction hypothesis, we know $\beta_1 \ge \sqrt{n_1}$ and
$\beta_2 \ge \sqrt{n_2}$.
We now proceed to calculate $\ADV_{\vec 1}(h)=\ADV_{(\beta_1,\beta_2)}(\AND)$.
We set up our AND adversary matrix as follows:
\begin{center}
\begin{tabular}{|c|cccc|}
\hline
   & 00 & 01 & 10 & 11 \\ \hline
00 & 0  & 0  & 0  & 0  \\
01 & 0  & 0  & 0  & $\beta_1$  \\
10 & 0  & 0  & 0  & $\beta_2$  \\
11 & 0  & $\beta_1$  & $\beta_2$  & 0  \\
\hline
\end{tabular}
\end{center}
This matrix has spectral norm $\sqrt{\beta_1^2 + \beta_2^2}$, and
$\sn{\Gamma \circ D_1}=\beta_1$, and
$\sn{\Gamma \circ D_2}=\beta_2$.  Thus
$$
\beta_1 \frac{\sn{\Gamma}}{\sn{\Gamma \circ D_1}}=
\beta_2 \frac{\sn{\Gamma}}{\sn{\Gamma \circ D_2}}=
\sqrt{\beta_1^2 + \beta_2^2}\ge \sqrt{n+1}.
$$
\end{proof}

\subsection{Composition Lemma}
\label{sec:comp_lemma}
We now turn to the proof of the composition theorem.
Given an adversary matrix $\Gf$ realizing the adversary bound for $f$
and adversary matrices $\Gg i$ realizing the adversary bound for
$g_i$ where $i=1, \ldots, k$, we build an adversary matrix $\Gh$
for the function $h=f \circ (g_1, \ldots, g_k)$.  \lemref{lem:gh} expresses
the spectral norm of this $\Gh$ in terms of the spectral norms of
$\Gf$ and $\Gg i$.   Moreover, if $\Gf,\Gg i$ are nonnegative, then
$\Gh$ will be nonnegative.

Let $\Gf$ be an adversary matrix for $f$, i.e.\ a Hermitian matrix satisfying
$\Gf[x,y]=0$ if
$f(x)=f(y)$, and let $\df$ be a prinicipal eigenvector of $\Gf$ with unit norm.
Similarly, let $\Gg{i}$ be a spectral matrix for
$g_i$ and let $\dg{i}$ be a principal eigenvector of unit norm, for every
$i=1, \ldots, k$.

It is helpful to visualize an adversary matrix in the following way.  Let
$X_{f}=f^{-1}(0)$ and $Y_{f}=f^{-1}(1)$.  We order the rows first by
elements from $X_{f}$ and then by elements of $Y_{f}$.  In this way, the
matrix has the following form:
$$
\Gf=
\left[
\begin{array}{cc}
0 & \Gf^{(0,1)}  \\
\Gf^{(1,0)} & 0   \\
\end{array}\right]
$$
where $\Gf^{(0,1)}$ is the submatrix of $\Gf$ with rows labeled
from $X_f$ and columns labeled from $Y_f$ and $\Gf^{(1,0)}$ is the
conjugate transpose of $\Gf^{(0,1)}$.

Thus one can see that an adversary matrix for a Boolean function corresponds
to a (weighted) bipartite graph
where the two color classes are the domains where the function takes
the values $0$ and $1$.  For
$b \in \01$ let $\dgb{i}{b}[x]=\dg{i}[x]$ if
$g_i(x)=b$ and $\dgb{i}{b}[x]=0$ otherwise.  In other words, $\dgb{i}{b}$ is
the vector $\dg{i}$ restricted to the color class $b$.

Before we define our composition matrix, we need one more piece of notation.
Let $\Gf^{(0,0)}=\sn{\Gf}I_{|X_f|}$, where $I$ is a $|X_f|$-by-$|X_f|$ identity
matrix and similarly $\Gf^{(1,1)}=\sn{\Gf}I_{|Y_f|}$.

We are now ready to define the matrix $\Gh$:
\begin{definition}
$
\Gh[x,y]=\Gf[\tx,\ty] \cdot
\left( \bigotimes_i \Gg{i}^{(\tx_i, \ty_i)} \right) [x,y]
$
\label{def:comp}
\end{definition}

\begin{lemma}
\label{lem:gh}
Let $\Gh$ be as in \defref{def:comp}.
Then $\sn{\Gh}=\sn{\Gf}\cdot \prod_{i=1}^k \sn{\Gg i}$ and a principal
eigenvector of $\Gh$ is
$\delta_h[x]=\df[\tx] \cdot \prod_{i=1}^k \dg{i}[x^i]$.
\label{lem:spnorm}
\end{lemma}

\begin{proof}
The more difficult direction is to show
$\sn{\Gh} \le \sn{\Gf}\cdot \prod_{i=1}^k \sn{\Gg i}$, and we do this first.
The outline of this direction is as follows:
\begin{enumerate}
  \item We first define $2^{k+n}$ many vectors
    $\delta_{\alpha,c} \in \C^{2^n}$.
  \item We show that each $\delta_{\alpha,c}$ is an eigenvector of
        $\Gh$.
  \item We show that $\{\delta_{\alpha,c}\}_{\alpha,c}$ span a space of
        dimension $2^n$. This implies that every eigenvalue of $\Gh$ is an
        eigenvalue associated to at least one of the $\delta_{\alpha,c}$ as
        eigenvectors corresponding to different eigenvalues of a
    symmetric matrix are orthogonal.
  \item We upper bound the absolute value of the eigenvalues corresponding to
    the $\delta_{\alpha,c}$ by $\sn{\Gf} \cdot \prod_{i=1}^k \sn{\Gg{i}}$.
\end{enumerate}

Let $c=(c_1, \ldots, c_k)$ where $c_i \in [2^{n_i}]$
for $i=1,\ldots,k$.  Let $\delta_{c_i}$ be an eigenvector of unit norm
corresponding to the \nth{c_i} largest eigenvalue of $\Gg{i}$---that is
$\Gg{i} \delta_{c_i} = \lambda_{c_i}(\Gg{i}) \delta_{c_i}$.

It is helpful to look at the matrix $\Gh$ as composed of blocks labeled
by $a,b \in \01^k$ where the $(a,b)$ block of the matrix consists of all
$x,y$ pairs with $\tx=a$ and $\ty=b$.  Notice that the $(a,b)$ block of
$\Gh$ is the matrix $\Gf[a,b] \cdot \otimes \Gg{i}^{(a_i, b_i)}$.

Let $\lambda_{c_i}^0(A)=\sn{A}$ and $\lambda_{c_i}^1(A)=\lambda_{c_i}(A)$.
We claim that
$\Gg{i}^{(a_i,b_i)} \dgc{i}{b_i}=
\lambda_{c_i}^{a_i \oplus b_i}(\Gg{i}) \dgc{i}{a_i}$.
This is because if $a_i \ne b_i$ then $\Gg{i}^{(a_i,b_i)}$ is one half of the
bipartite matrix $\Gg{i}$ and so
$\Gg{i}^{(a_i,b_i)} \dgc{i}{b_i}=\lambda_{c_i}(\Gg{i}) \dgc{i}{a_i}$.
On the other hand, if $a_i=b_i$ then $\Gg{i}^{(a_i,b_i)}=\sn{\Gg{i}}I$ and so
$\Gg{i}^{(a_i,b_i)} \dgc{i}{b_i}=\sn{\Gg{i}} \dgc{i}{b_i}=
\sn{\Gg{i}} \dgc{i}{a_i}$.

Thus for the tensor product matrix $\otimes \Gg{i}^{(a_i, b_i)}$ we have that
$$
\otimes \Gg{i}^{(a_i, b_i)} \otimes \dgc{i}{b_i}=
\prod_{i=1}^k \lambda_{c_i}^{a_i \oplus b_i}(\Gg{i}) \cdot \otimes \dgc{i}{a_i}.
$$
Expanding this equation gives that for every $x$ such that $\tx=a$
\begin{equation}
\sum_{y:\ty=b} \otimes \Gg{i}^{(a_i,b_i)}[x,y] \cdot
(\otimes \delta_{c_i})[y] =
\prod_{i=1}^k \lambda_{c_i}^{a_i \oplus b_i}(\Gg{i}) \cdot
(\otimes \delta_{c_i})[x].
\label{eigeq1}
\end{equation}

Now consider a $2^k$-by-$2^k$ matrix $A_c$ where
$$
A_c[a,b]=\Gf[a,b] \cdot \prod_{i=1}^k \lambda_{c_i}^{a_i \oplus b_i}(\Gg{i}).
$$
Let $\alpha$ be a unit norm eigenvector of this matrix, say with eigenvalue
$\mu_{\alpha,c}$.  Explicitly writing out the eigenvalue equation means
that for every $a$,
\begin{equation}
\sum_b \Gf[a,b]
\cdot \prod_{i=1}^k \lambda_{c_i}^{a_i \oplus b_i}(\Gg{i}) \cdot \alpha[b]=
\mu_{\alpha,c} \; \alpha[a].
\label{eigeq2}
\end{equation}

\paragraph{Item 1:}
We are ready to define our proposed eigenvectors of $\Gh$.
For any $c=(c_1, \ldots, c_k)$ and $\alpha$ an eigenvector of $A_c$
let
$$
\delta_{\alpha,c}[x]=\alpha[\tx]\cdot \prod_{i=1}^k \delta_{c_i} [x^i]
=\alpha[\tx] \cdot (\otimes \delta_{c_i}) [x].
$$

\paragraph{Item 2:}
We claim that $\delta_{\alpha,c}$ is an eigenvector of $\Gh$ with
eigenvalue $\mu_{\alpha, c}$.
This can be verified as follows: for any $x$,
\begin{align*}
\sum_y \Gh[x,y] \delta_{\alpha,c}[y] &=
\sum_y \Gf[\tx,\ty] \alpha[\ty] \cdot (\otimes \Gg{i}^{(\tx_i, \ty_i)})[x,y]
\cdot (\otimes \delta_{c_i})[y] \\
&= \sum_b \Gf[\tx,b] \alpha[b] \cdot
\sum_{y:\ty=b} (\otimes \Gg{i}^{(\tx_i, \ty_i)})[x,y]
\cdot (\otimes \delta_{c_i})[y] \\
\end{align*}

Applying \eqnref{eigeq1} gives
\begin{align*}
\sum_y \Gh[x,y] \delta_{\alpha,c}[y] &=
\sum_b \Gf[\tx,b] \alpha[b] \cdot
\prod_{i=1}^k \lambda_{c_i}^{\tx_i \oplus b_i}(\Gg{i}) \cdot
(\otimes \delta_{c_i})[x] \\
&= (\otimes \delta_{c_i})[x] \cdot \sum_b \Gf[\tx,b]
\cdot \prod_{i=1}^k \lambda_{c_i}^{\tx_i \oplus b_i}(\Gg{i}) \alpha[b].
\end{align*}

And now applying \eqnref{eigeq2} gives
\begin{equation*}
\sum_y \Gh[x,y] \delta_{\alpha,c}[y] = \mu_{\alpha,c} \alpha[\tx] \cdot
(\otimes \delta_{c_i})[x]
= \mu_{\alpha,c} \; \delta_{\alpha,c}[x].
\end{equation*}
Thus $\delta_{\alpha,c}$ is an eigenvector of $\Gh$ with eigenvalue
$\mu_{\alpha,c}$.  This completes the second step of the proof.

\paragraph{Item 3:}
We now claim that the vectors $\{\delta_{\alpha,c}\}_{\alpha,c}$ span
$\C^{2^n}$.  For a fixed $c$, the set
of eigenvectors $\{ \alpha_\ell \}_{\ell=1}^{2^k}$ of $A_c$ forms an
orthogonal basis for the space of vectors of dimension $2^k$, hence there is a
linear combination $\gamma$ of $\alpha_\ell$'s such that $\sum_\ell
\gamma_\ell \alpha_\ell = (1, 1, \dots, 1)$.  Then $\sum_\ell \gamma_\ell
\delta_{\alpha_\ell,c} = \otimes \delta_{c_i}$.  Now, since
$\{ \delta_{c_i} \}_{c_i=1}^{2^{n_i}}$ form an orthogonal basis for every $i$,
linear combinations of $\delta_{\alpha,c}$ span the whole space of dimension
$2^{\sum_i n_i}$, which is the dimension of $\Gh$.  Hence every eigenvector
of $\Gh$ can be expressed in this form.  This completes step three of the
proof.

\paragraph{Item 4:}
It now remains to show that
$\mu_{\alpha,c} \le \sn{\Gf} \cdot \prod_i \sn{\Gg{i}}$ for every $\alpha,c$.
To do this, fix $c$ and consider the matrix $A_c$.
\begin{equation}
\label{eq:mu}
\mu_{\alpha,c} = \alpha^* A_c \alpha
= \sum_{a,b} \Gf[a,b] \cdot \prod_{i=1}^k
\lambda_{c_i}^{a_i \oplus b_i} (\Gg i) \cdot
\alpha[a] \alpha[b].
\end{equation}

Notice that $-\sn{\Gg{i}} \le \lambda_{c_i}(\Gg{i}) \le \sn{\Gg{i}}$.  Our
first claim is that we can replace
$\lambda_{c_i}(\Gg{i})$ by either $\sn{\Gg{i}}$ or $-\sn{\Gg{i}}$ in such
a way that the sum in \eqref{eq:mu} does not decrease.
To see this, we can first factor out $\lambda_{c_1}(\Gg{1})$ of the above sum
and look at the term it multiplies.  If this term is positive, then setting
$\lambda_{c_1}(\Gg{1})$ to $\sn{\Gg{1}}$ will not decrease the sum; on the
other hand, if the term it multiplies is negative, then replacing
$\lambda_{c_1}(\Gg{1})$ by $-\sn{\Gg{1}}$ will not decrease the sum.  We
continue this process in turn with $i=2,\ldots, k$.

Let $d_i=1$ if in this process we replaced $\lambda_{c_i}(\Gg{i})$ by
$-\sn{\Gg{i}}$ and $d_i=0$ if $\lambda_{c_i}(\Gg{i})$ was replaced by
$\sn{\Gg{i}}$.  Note that if $a_i=b_i$, then no replacement was made and
the coefficient remains $\sn{\Gg{i}}$.  We thus now have
\begin{equation}
\mu_{\alpha,c} \le
\sum_{a,b} \Gf[a,b] \alpha[a] \alpha[b] \cdot
\prod_{i=1}^k (-1)^{d_i (a_i + b_i)} \sn{\Gg{i}},
\label{eq:replacement}
\end{equation}
A key fact here is that the sign
of $\sn{\Gg{i}}$ will be the same everywhere $a_i \ne b_i$---the
signs of entries cannot be flipped at will.

We now mimic the pattern of signs in \eqnref{eq:replacement} by defining a
new unit vector $\alpha'$.  Let
$\alpha'[a]=\alpha[a] \prod_i (-1)^{d_i \cdot a_i}$.
Then
\begin{align*}
\mu_{\alpha,c} &\le
\sum_{a,b} \Gf[a,b] \alpha[a] \alpha[b] \cdot
\prod_{i=1}^k (-1)^{d_i (a_i + b_i)} \sn{\Gg{i}} \\
&= \prod_{i=1}^k \sn{\Gg{i}} \sum_{a,b} \Gf[a,b] \alpha'[a] \alpha'[b] \\
&\le \sn{\Gf} \cdot \prod \sn{\Gg{i}},
\end{align*}
which we wished to show.

\paragraph{Other direction:}
We now show that $\sn{\Gh} \ge \sn{\Gf} \cdot \prod_{i=1}^k \sn{\Gg{i}}$.
Let $\df$ be a principal eigenvector of $\Gf$ and $\dg{i}$ a principal
eigenvector for $\Gg{i}$ for $i=1, \ldots, k$.  We have already argued that
$\delta_h=\df[\tx] \cdot \prod_{i=1}^k \dg{i}[x^i]$ is an eigenvector
of $\Gh$ whose eigenvalue is the eigenvalue of the matrix $A_{\vec 1}$
where
$$
A_{\vec 1}[a,b] = \Gf[a,b] \cdot \prod_{i=1}^k \sn{\Gg{i}}.
$$
Factoring out $\prod_{i=1}^k \sn{\Gg{i}}$ from $A_{\vec 1}$ we are simply left
with the matrix $\Gf$, thus the largest eigenvalue of $A_{\vec 1}$ is
$\sn{\Gf} \cdot \prod_{i=1}^k \sn{\Gg{i}}$.
\end{proof}

\subsection{Composition lower bound}
\label{sec:lower}
With \lemref{lem:spnorm} in hand, it is a relatively easy matter to show
a lower bound on the adversary value of the composed function $h$.
Let $\BADV$ denote either $\ADV$ or $\MADV$.

\begin{lemma}
  \label{lem:compose-sa}
$\BADV_\alpha(h) \ge \BADV_\beta(f)$,
where $\beta_i = \BADV_{\alpha^i}(g_i)$,
\end{lemma}

\begin{proof}
Due to the maximization over all matrices $\Gamma$, the spectral bound of the
composite function $h$ is at least $\BADV_\alpha(h) \ge \min_{\ell=1}^n (
\alpha_\ell \sn{\Gh} / \sublam \Gh \ell )$, where $\Gh$ is defined as in
\lemref{lem:spnorm}.  Notice that in $\lemref{lem:spnorm}$, if the component
matrices are nonnegative, then $Gh$ will be as well, thus we can
simultaneously treat both adversary bounds.

We compute $\sublam \Gh \ell$ for $\ell = 1, \dots, n$.
Let the \nth{\ell} input bit be the \nth{q} bit in the \nth{p} block.
Recall that
\begin{align*}
\Gh[x,y]
  &= \Gf[\tx,\ty] \cdot \prod_{i=1}^k \Gg{i}^{(\tx_i, \ty_i)}[x^i,y^i]. \\
\noalign{We prove that} \\
(\Gh \circ D_\ell)[x,y]
  &= (\Gf \circ D_p) [\tx,\ty]
  \cdot (\Gg{p}\circ D_q)^{(\tx_p, \ty_p)}[x^p,y^p]
  \cdot \prod_{i \ne p} \Gg{i}^{(\tx_i, \ty_i)} [x^i,y^i].
\end{align*}

If $x_\ell \ne y_\ell$ and $\tx_p \ne \ty_p$ then both sides are equal because
all multiplications by $D_p,D_q,D_\ell$ are multiplications by 1.  If this
is not the case---that is, if $x_\ell = y_\ell$ or $\tx_p = \ty_p$---then both
sides are zero.  We see this by means of two cases:

\begin{enumerate}
  \item $x_\ell = y_\ell$: In this case the left hand side is zero due to
  $(\Gh \circ D_\ell) [x,y] = 0$.  The right hand side is also zero because
    \begin{enumerate}
      \item if $\tx_p = \ty_p$ then the right hand side is zero as
      $(\Gf \circ D_p) [\tx,\ty] = 0$.
      \item else if $\tx_p \ne \ty_p$ then the right hand side is zero as
      $(\Gg p \circ D_q) [x^p, y^p] = 0$.
    \end{enumerate}
  \item $x_\ell \ne y_\ell$, $\tx_p = \ty_p$: The left side is zero because
    $\Gg{p}^{(\tx_p, \ty_p)}[x^p,y^p]=
    \sn{\Gg{p}}I[x^p,y^p]=0$ since $x^p \ne y^p$.
    The right side is also zero due to $(\Gf \circ D_p) [\tx,\ty] = 0$.
\end{enumerate}

Since $\Gh \circ D_\ell$ has the same structure as $\Gh$, by
\lemref{lem:spnorm}, $\sublam \Gh \ell = \sublam \Gf p \cdot \sublam
{\Gg p} q \cdot \prod_{i \ne p} \sn{\Gg i}$.  By dividing the two
spectral norms,
\begin{equation}
{\sn{\Gh} \over \sublam \Gh \ell}
  = {\sn{\Gf} \over \sublam \Gf p}
  \cdot {\sn{\Gg p} \over \sublam {\Gg p} q}.
\label{eq:lg/lgi}
\end{equation}

Since the spectral adversary maximizes over all adversary matrices, we
conclude that
\begin{align*}
\BADV_\alpha(h)
&\ge \min_{\ell=1}^n {\sn{\Gh} \over \sublam \Gh \ell} \cdot \alpha_\ell \\
&= \min_{i=1}^k {\sn{\Gf} \over \sublam \Gf i}
  \cdot \min_{j=1}^{n_i} {\sn{\Gg i} \over \sublam {\Gg i} {j}}
  \cdot \alpha^i_j \\
&= \min_{i=1}^k {\sn{\Gf} \over \sublam \Gf i}
  \cdot \BADV_{\alpha^i}(g_i) \\
&= \min_{i=1}^k {\sn{\Gf} \over \sublam \Gf i } \cdot \beta_i \\
&= \BADV_\beta(f),
\end{align*}
which we had to prove.
\end{proof}

\subsection{Composition upper bound}
\label{sec:upper}

The non-negative adversary bound $\ADV$ satisfies the matching upper bound
$\ADV_\alpha(h) \le \ADV_\beta(f)$.
Interestingly, we do not know yet how to show this for the $\MADV$ bound.

We apply the duality theory of semidefinite programming to obtain an
equivalent expression for $\ADV_\alpha$ in terms of a minimization problem.
We then upper bound $\ADV_\alpha(h)$ by showing how to compose solutions
to the minimization problems.

\begin{definition}
Let $f: S \to \01$ be a partial boolean function, where $S \subseteq
\01^n$, and let $\alpha \in \R_{+}^n$.
The \emph{minimax bound of $f$ with costs $\alpha$} is
\[
\MM_\alpha(f) = \min_p \max_{\fxnefy f}
  \frac{1}{\sum_{\xineyi i} \sqrt{ p_x(i) p_y(i) } / \alpha_i},
\]
where $p: S \times [n] \to [0,1]$ ranges over all sets of $|S|$
probability distributions over input bits, that is, $p_x(i) \ge 0$ and
$\sum_i p_x(i) = 1$ for every $x \in S$.
\end{definition}

This definition is a natural generalization of the minimax bound introduced
in \cite{LM04, SS06}.  As \cite{SS06} show that the minimax bound is equal to
the spectral norm formulation of the adversary method, one can similarly show
that the versions of these methods with costs are equal.

\begin{theorem}
[Duality of adversary bounds]
\label{thm:equal}
For every $f: \01^n \to \01$ and $\alpha \in \R_{+}^n$,
\begin{equation*}
\ADV_\alpha(f) = \MM_\alpha(f).
\end{equation*}
\end{theorem}

\begin{proof}[Sketch of proof.]
We start with the minimax bound with costs, substitute $q_x(i) p_x(i) / \alpha_i$, and rewrite the condition $\sum_i p_x(i) = 1$ into
$\sum_i \alpha_i q_x(i) = 1$.  Using similar arguments as
in~\cite{SS06}, we rewrite the bound as a semidefinite program,
compute its dual, and after a few simplifications, get the spectral
bound with costs.
\end{proof}

\begin{lemma}
  \label{lem:compose-mm}
$\ADV_\alpha(h) \le \ADV_\beta(f)$.
\end{lemma}

\begin{proof}
Let $p^f$ and $p^{g_i}$ for $i=1, \dots, k$ be optimal sets of probability
distributions achieving the minimax bounds.  Thus using \thmref{thm:equal}
we have
\begin{align*}
\ADV_\beta(f) &= \max_{\fxnefy f} {1 \over \sum_{\xineyi i}
    \sqrt{ p^f_x(i) p^f_y(i) } / \beta_i }, \\
\ADV_{\alpha^i}(g_i) &= \max_{\fxnefy {g_i}} {1 \over \sum_{\xineyi j}
    \sqrt{ p^{g_i}_x(j) p^{g_i}_y(j) } / \alpha^i_j}. \\
\end{align*}
Define the set of probability distributions $p^h$ as
$p^h_x(\ell)= p^f_\tx(i) p^{g_i}_{x^i}(j)$, where the
\nth{\ell} input bit is the \nth{j} bit in the \nth{i} block.  This
construction was used by
Laplante, Lee, and Szegedy~\cite{LLS06} to prove a similar
bound in the stronger setting where the sub-functions $g_i$ can act on
the same input bits.
%\footnote{This setting is, however, not applicable to us, because
%we cannot prove a matching lower bound for $\ADV_\alpha(h)$.}
We claim that $p^h$ witnesses that $\ADV_\alpha(h) \le \ADV_\beta(f)$:
\begin{align*}
\ADV_\alpha(h)
&\le \max_{\fxnefy h} {1 \over \sum_{\xineyi \ell} \sqrt{ p^h_x(\ell)
p^h_y(\ell) } / \alpha_\ell } \\
&= 1 \Bigg/ \min_{\fxnefy h} \sum_{\xineyi \ell} \sqrt{ p^f_\tx(i) p^f_\ty(i) }
    \sqrt{ p^{g_i}_{x^i}(j) p^{g_i}_{y^i}(j) }
    / \alpha^i_j \\
&= 1 \Bigg/ \min_{\tx, \ty \atop f(\tx) \ne f(\ty)}
    \sum_i \sqrt{ p^f_\tx(i) p^f_\ty(i) }
    \min_{x^i, y^i \atop {g_i(x^i)=\tx_i \atop g_i(y^i)=\ty_i}}
    \sum_{j: x^i_j \ne y^i_j} \sqrt{ p^{g_i}_{x^i}(j) p^{g_i}_{y^i}(j) }
    / \alpha^i_j \\
&\le 1 \Bigg/ \min_{\tx, \ty \atop f(\tx) \ne f(\ty)}
    \sum_{i: \tx_i \ne \ty_i} \sqrt{ p^f_\tx(i) p^f_\ty(i) }
    \min_{x^i, y^i \atop g_i(x^i) \ne g_i(y^i)}
    \sum_{j: x^i_j \ne y^i_j} \sqrt{ p^{g_i}_{x^i}(j) p^{g_i}_{y^i}(j) }
    / \alpha^i_j \\
&= 1 \Bigg/ \min_{\tx, \ty \atop f(\tx) \ne f(\ty)}
    \sum_{i: \tx_i \ne \ty_i} \sqrt{ p^f_\tx(i) p^f_\ty(i) }
    \ /\  \ADV_{\alpha^i}(g_i) \\
&= 1 \Bigg/ \min_{\tx, \ty \atop f(\tx) \ne f(\ty)}
    \sum_{i: \tx_i \ne \ty_i} \sqrt{ p^f_\tx(i) p^f_\ty(i) }
    \ /\  \beta_i \\
&= \ADV_\beta(f),
\end{align*}
where the second inequality follows from that fact that we have
removed $i: \tx_i = \ty_i$ from the sum and the last equality follows from
\thmref{thm:equal}.
\end{proof}

\section{Examples}
\label{sec:example}
In this section, we look at some examples to see how negative weights can
help to achieve larger lower bounds.  We consider two examples in detail: a
4-bit function giving the largest known separation between the polynomial
degree and quantum query complexity, and a 6-bit function breaking the
certificate complexity and property testing barriers.

To help find good adversary matrices, we implemented both adversary bounds
as semidefinite programs and used the convex optimization package SeDuMi
\cite{sedumi} for Matlab.  Using these programs, we tested both $\ADV$ and
$\MADV$ bounds for
all 222 functions on 4 or fewer variables which are not equivalent under
negation of output and input variables and permutation of input variables
(see sequence number A000370 in \cite{Slo}).
The $\MADV$ bound is strictly larger than the $\ADV$ bound for 128 of these
functions.  The source code of our semidefinite programs and more examples can
be downloaded from~\cite{code}.

\subsection{Ambainis function}
In order to separate quantum query complexity and polynomial degree,
Ambainis defines a Boolean function $f: \01^4 \to \01$ which is one if and
only if the four input bits are sorted%
\footnote{The function was first described in this
way by Laplante, Lee, and Szegedy \cite{LLS06}.  The function defined by
Ambainis \cite{Amb03} can be obtained from this function by exchanging the
first and third input bits and negating the output.}, that is they are either
in a non-increasing or non-decreasing order.  This function has polynomial
degree 2, and an adversary bound of 2.5.  Thus by the composition theorem for
the nonnegative adversary method, Ambainis obtains a separation between quantum
query complexity and polynomial
degree of $Q_{\epsilon}(f^d)=\Omega(\deg(f^d)^{1.321})$.
We have verified that this function indeed gives the largest separation between
adversary bounds and polynomial degree over all functions on 4 or fewer
variables.

In the next theorem, we construct an adversary matrix with negative weights
which shows that $\MADV(f) \ge 2.5135$.  Using the composition theorem
\thmref{thm:madvcomposition} we obtain $\MADV(f) \ge \ADV(f)^{1.005}$ and
improve the separation between quantum query complexity and polynomial degree
to $Q_{\epsilon}(f^d)=\Omega(\deg(f^d)^{1.3296})$.

\begin{theorem}
Let $f:\01^4 \rightarrow \01$ be Ambainis' function.  Then
$\MADV(f) \ge 2.5135$.
\end{theorem}

\begin{proof}
Following our own advice, to design a good adversary
matrix the first thing we do is look at the automorphisms of the function.
Notice that the element $g=(1432) \times ((01),\id,\id,\id) \in S_4 \times
S_{\01}^4$ preserves the property of the bits being ordered, and thus
also the function value.  We are using cycle notation here, so for example,
$(1432) \times ((01),\id,\id,\id) \cdot 0000=0001$.  Let $G$ be the group
generated by $g$.  As $g$ has order eight, this group is isomorphic to
$\mathbb{Z}_8$.  The group $G$ is both $f$-transitive and index transitive,
thus we know that the
uniform vector will be a principal eigenvector of our eventual adversary
matrix, and that all $\Gamma \circ D_i$ will be unitarily equivalent.

We now construct our adversary matrix.  If $f(x)=f(y)$ we set $\Gamma[x,y]=0$.
So now we just consider the ``interesting'' part of the adversary matrix with
rows labeled by inputs which map to one, and columns labeled by inputs which
map to 0.  To highlight the group structure in the matrix, we let the
$i^{th}$ row be $g^i(0000)$ and similarly let the $i^{th}$ column be
$g^i(0010)$.

It turns out there are four types of pairs $(x,y)$ which are not equivalent
under automorphism. We let $\Gamma[x,y]=a$ if (x,y) have Hamming distance one.
If $(x,y)$ have Hamming distance 3, they are also related by automorphism, and
we set $\Gamma[x,y]=d$ in this case.  There are two automorphism types for
$(x,y)$ pairs which have Hamming distance 2.  If they differ on bits which
are both sensitive or both not sensitive, we set $\Gamma[x,y]=b$; otherwise,
we set $\Gamma[x,y]=c$.  The adversary matrix then looks as follows:

$$
\begin{tabular}{|c|cccccccc|}
\hline
      & 0010 & 0101 & 1011 & 0110 & 1101 & 1010 & 0100 & 1001 \\ \hline
0000  &  a   &  c   &  d   &  b   &  d   &  c   &  a   & b    \\
0001  &  b   &  a   &  c   &  d   &  b   &  d   &  c   & a    \\
0011  &  a   &  b   &  a   &  c   &  d   &  b   &  d   & c    \\
0111  &  c   &  a   &  b   &  a   &  c   &  d   &  b   & d    \\
1111  & \bf d& \bf c&  a   & \bf b&  a   &  c   & \bf d& b    \\
1110  & \bf b& \bf d&  c   & \bf a&  b   &  a   & \bf c& d    \\
1100  & \bf d& \bf b&  d   & \bf c&  a   &  b   & \bf a& c    \\
1000  & \bf c& \bf d&  b   & \bf d&  c   &  a   & \bf b& a    \\
\hline
\end{tabular}
$$

As we have remarked, all $\Gamma \circ D_i$ matrices are equivalent up to
permutation, and it can be shown that they
consist of four $4$-by-$4$ disjoint blocks, each of these blocks being some
permutation of rows and columns of the following matrix $B$:
\begin{equation}
B = \left( \begin{matrix}
c & b & d & d \\
b & c & d & a \\
d & d & c & b \\
d & a & b & c \\
\end{matrix} \right).
\end{equation}

The particular block $B$ above is one of the four blocks of $\Gamma \circ
D_1$ with columns indexed by zero-inputs 0010, 0100, 0101, 0110, and rows
indexed by one-inputs 1000, 1110, 1111, 1100.  A
principal eigenvector of $\Gamma$ is given by the uniform vector, which
has eigenvalue $2(a+b+c+d)$.  Thus our optimization problem is to maximize
$2(a+b+c+d)$ while keeping $\sn{B}\le 1$.  The optimal
setting of the four variables can be found numerically by semidefinite
programming and is the following:
\begin{equation}
\begin{array}{| r | r r |}
\hline
& \ADV & \MADV \\
\hline
a & 3/4 & 0.5788 \\
b & 1/2 & 0.7065 \\
c & 0   & 0.1834 \\
d & 0   &-0.2120 \\
\lambda & 5/2 & 2.5135 \\
\hline
\end{array}
\end{equation}

The eigenvalues of $\Gamma \circ D_i$ are $\{ 1, 1, \frac14, \frac14 \}$,
and the eigenvalues of $\Gamma^{\pm} \circ D_i$ are $\{ 1, 1, -1,
-0.2664 \}$.  Both spectral bounds are tight due to the existence
of matching dual solutions; we, however, omit them here.
\end{proof}

\subsection{Breaking the certificate complexity barrier}
We now consider a function on six bits.  We will consider this function in
two guises.  We first define a partial function $f$ to show that $\MADV$
can break the property testing barrier.  We then extend this partial function
to a total monotone function $g$ which gives a larger separation
between the $\ADV$ and $\MADV$ bounds, and also shows that $\MADV$ can
break the certificate complexity barrier.

We define the partial function $f$ on six bits as follows:
\begin{itemize}
 \item The zero inputs of $f$ are:
111000, 011100, 001110, 100110, 110010, 101001, 010101, 001011, 100101, 010011.
 \item The one inputs of $f$ are: 000111, 100011, 110001, 011001,
001101, 010110, 101010, 110100, 011010, 101100.
\end{itemize}
Notice that $f$ is defined on all inputs with Hamming weight three, and only on these inputs.  This function is inspired by a function defined by Kushilevitz
which appears in \cite{NW95} and is also discussed by Ambainis \cite{Amb03}.
Kushilevitz's function has the same behavior as the
above on inputs of Hamming weight three; it is additionally defined to be 0 on
inputs with Hamming weight 0, 4, or 5, and to be 1 on inputs with Hamming
weight 1, 2, or 6.

All zero inputs of $f$ have Hamming distance at least 2 from any
one input, thus the relative Hamming distance between any zero and one
input is $\epsilon=1/3$.  In \thmref{thm:sixbit} we show that
$\MADV(f) \ge 2 +3\sqrt{5}/5 \approx 3.341$.  This implies
$\MADV(f) \ge (1/\epsilon(f))^{1.098}$, and as both bounds compose we obtain
$\MADV(f^d) \ge (1/\epsilon(f^d))^{1.098}$.
This shows that the property testing barrier does not apply to $\MADV$
as it does to $\ADV$.  The relative Hamming distance $\epsilon(f^d)$, however,
goes to zero when $d$ increases.  We don't know of an asymptotic separation for
constant $\epsilon$.

We also consider a monotone extension of $f$ to a total function, denoted $g$.
It is additionally defined to be 0 on inputs with Hamming weight 0, 1, or 2,
and to be 1 on inputs with Hamming weight 4, 5, or 6.
Recall that the maxterms of a monotone Boolean function are the maximal,
under subset ordering, inputs $x$ which evaluate to 0, and similarly the
minterms are the minimal inputs which evaluate to 1.  The zero inputs of $f$
become maxterms of $g$ and the one inputs become minterms.  Since $f$ is
defined on all inputs
with Hamming weight three, $g$ is a total function.  The extended function
$g$ is at least as hard as its sub-function $f$, hence $\MADV(g) \ge \MADV(f)$.  The
0-certificates of $g$ are given by the location of 0's in the maxterms and the
1-certificates are given by the location of 1's in the minterms, thus
$C_0(g)=C_1(g)=3$.  Both bounds compose thus $C_0(g^d)=C_1(g^d)=3^d$.

Applying the composition theorem
\thmref{thm:madvcomposition} we obtain
$\MADV(g^d) \ge (C_0(g)C_1(g))^{0.549}$.  As
$\ADV(h) \le \sqrt{C_0(h)C_1(h)}$ for a total function $h$, we also conclude
$\MADV(g^d) \ge \ADV(g^d)^{1.098}$.

\begin{theorem}
\label{thm:sixbit}
$\MADV(f) \ge 2+3\sqrt{5}/5$.
\end{theorem}

\begin{proof}
To design an adversary matrix for $f$, we again first consider its
automorphisms.  The automorphism group $G$ of $f$ contains
a subgroup isomorphic to $A_5$, the alternating group on five elements.
As we have listed the zero and one instances of the function, one can easily
see that the permutation $(12345)$, in cycle notation, is an automorphism.
This automorphism fixes the
sixth bit.  It turns out that for every $1 \le i \le 6$ there
is an automorphism of $f$ of order 5 which fixes the $i^{th}$ bit. Here are
some examples: $(25643),(15643),(12456),(16235),(14362),
(12345)$.  Taking the closure of these elements gives a group isomorphic to
$A_5$.  This group is $f$-transitive and index transitive, thus we know that
the uniform eigenvector will be a principal eigenvector of our eventual
adversary matrix, and that all $\Gamma \circ D_i$ will have the same
spectral norm.

Any two pairs $(x,y)$ and $(x',y')$ with the same Hamming distance are
related by automorphism, thus the $(x,y)$ entry of $\Gamma$ will only depend
on Hamming distance.  As all valid inputs to $f$ have Hamming weight three,
the Hamming distance between $x$ and $y$ is even and is either two, four, or
six.  We label the matrix entries $a,b,c$ respectively for Hamming distances
two, four, six.

\[
\begin{tabular}{|c|c c c c c|c c c c c|}
\hline
      &000&100&110&011&001&010&101&110&011&101
\\    &111&011&001&001&101&110&010&100&010&100
\\ \hline
111000& c    &  b   &  a   &  a   &  b   &  b  &  a   &  a   &  a   &  a
\\
011100& b    &  c   &  b   &  a   &  a   &  a   &  b   &  a  &  a   &  a
\\
001110& a    &  b   &  c   &  b   &  a   &  a   &  a   &  b  &  a   &  a
\\
100110& a    &  a   &  b   &  c   &  b   &  a   &  a   &  a  &  b   &  a
\\
110010& b    &  a   &  a   &  b   &  c   &  a   &  a   &  a  &  a   &  b
\\ \hline
101001& b    &  a   &  a   &  a   &  a   &  c   &  a   &  b  &  b   &  a
\\
010101& a    &  b   &  a   &  a   &  a   &  a   &  c   &  a  &  b   &  b
\\
001011& a    &  a   &  b   &  a   &  a   &  b   &  a   &  c  &  a   &  b
\\
100101& a    &  a   &  a   &  b   &  a   &  b   &  b   &  a  &  c   &  a
\\
010011& a    &  a   &  a   &  a   &  b   &  a   &  b   &  b  &  a   &  c
\\ \hline
\end{tabular}
\]
Notice that all rows have the same sum, thus the uniform vector is an
eigenvector with eigenvalue $6a+3b+c$.

From this ordering of rows and columns, one can easily read off the matrix
$\Gamma \circ D_6$. This is a block diagonal matrix with each block
equal, up to permutation, to a matrix $B$:
\[
B=\left(\begin{matrix}
c&b&a&a&b \\
b&c&b&a&a \\
a&b&c&b&a \\
a&a&b&c&b \\
b&a&a&b&c \\
\end{matrix}\right).
\]
Our optimization problem then becomes: maximize $6a+3b+c$ while keeping
the spectral norm of $B$ at most one.  As $B$ is a sum of circulants, its
eigenvalues will be $c+b\omega^k+a\omega^{2k}+a\omega^{3k}+b\omega^{4k}$,
for $0\le k \le 4$, where $\omega$ is a $5^{th}$ root of unity.  An optimal
solution is obtained by setting $a=(1+\sqrt{5})/5, b=(1-\sqrt{5})/5,c=1/5$.
This makes the eigenvalues of $B$ equal to $\{ 1,1,1,-1,-1\}$,
while $6a+3b+c=2+3\sqrt{5}/5$.
\end{proof}

\section{Conclusion}
Breaking the certificate complexity and property testing barriers opens the
possibility that $\MADV$ can prove better lower bounds where we know $\ADV$
cannot.  Salient examples are element distinctness, the collision problem,
and triangle finding.  For element distinctness, the best bound provable by
the standard adversary method is $O(\sqrt{n})$ while the polynomial method
is able to prove a tight lower bound of $\Omega(n^{2/3})$ \cite{AS04}. For
the collision problem, the adversary method is only able to prove a constant
lower bound while the polynomial method again proves a tight lower bound of
$\Omega(n^{1/3})$ \cite{AS04}.  Finally, for the problem of determining if a
graph contains a triangle, the best bound provable by the adversary method is
$O(n)$ and the best known algorithm is $O(n^{1.3})$ \cite{MSS05}.

It is also interesting to determine what limitations our new adversary method
might face.  The only limitation we are aware of is that the square of
$\MADV$ is a lower bound on formula size.  This is probably not a limitation,
however, as \cite{FGG07,CRSZ07} have recently taken a major step towards
proving the conjecture of Laplante, Lee, and Szegedy
that the square of bounded-error quantum query complexity is in general a lower
bound on formula size.

\section*{Acknowledgements}
We would like to thank Aram Harrow and Umesh Vazirani for interesting
discussions on the topics of this paper, and Ronald de Wolf for many
valuable comments on an earlier draft.

%\bibliographystyle{alpha}
%\bibliography{qc}
\newcommand{\etalchar}[1]{$^{#1}$}
 \urlstyle{tt} \newcommand{\quot}{\"} \newcommand{\leftbrace}{\{}
  \newcommand{\rightbrace}{\}}

\end{document}